\documentclass[acmsmall]{acmart}

\AtBeginDocument{%
  \providecommand\BibTeX{{%
    \normalfont B\kern-0.5em{\scshape i\kern-0.25em b}\kern-0.8em\TeX}}}

\setcopyright{rightsretained}
\acmJournal{PACMHCI}
\acmYear{2021} \acmVolume{5} \acmNumber{CSCW1} \acmArticle{18} \acmMonth{4} \acmPrice{}\acmDOI{10.1145/3449092}

\usepackage{booktabs}
\usepackage{subcaption, multirow}
\usepackage{color, colortbl}
\usepackage{graphicx}
\usepackage{longtable}
\usepackage{pifont}
\usepackage{titlesec}
\usepackage{amsmath}
\usepackage{xcolor}
\usepackage{afterpage} 
\begin{document}

\title{Exploring Lightweight Interventions at Posting Time to Reduce the Sharing of Misinformation on Social Media}


\author{Farnaz Jahanbakhsh}
\affiliation{%
  \institution{Computer Science and Artificial Intelligence Laboratory, Massachusetts Institute of Technology}
  \city{Cambridge}
  \country{USA}
}

\author{Amy X. Zhang}
\affiliation{%
  \institution{Allen School of Computer Science \& Engineering, University of Washington}
  \city{Seattle}
  \country{USA}
}

\author{Adam J. Berinsky}
\affiliation{%
  \institution{Political Science, Massachusetts Institute of Technology}
  \city{Cambridge}
  \country{USA}
}

\author{Gordon Pennycook}
\affiliation{%
  \institution{Hill/Levene Schools of Business, University of Regina}
  \city{Regina}
  \country{Canada}
}

\author{David G. Rand}
\affiliation{%
  \institution{Sloan School of Management/Brain and Cognitive Sciences, Massachusetts Institute of Technology}
  \city{Cambridge}
  \country{USA}
}

\author{David R. Karger}
\affiliation{%
  \institution{Computer Science and Artificial Intelligence Laboratory, Massachusetts Institute of Technology}
  \city{Cambridge}
  \country{USA}
}

\renewcommand{\shortauthors}{Farnaz Jahanbakhsh et al.}

\begin{abstract}

When users on social media share content without considering its veracity, they may unwittingly be spreading misinformation.
In this work, we investigate the design of lightweight interventions that nudge users to assess the accuracy of information as they share it. Such assessment may deter users from posting misinformation in the first place, and their assessments may also provide useful guidance to friends aiming to assess those posts themselves.

In support of lightweight assessment, we first develop a taxonomy of the reasons why people believe a news claim is or is not true; this taxonomy yields a checklist that can be used at posting time. 
We conduct evaluations to demonstrate that the checklist is an accurate and comprehensive encapsulation of people's free-response rationales.

In a second experiment, we study the effects of three behavioral nudges---1) checkboxes indicating whether headings are accurate, 2) tagging reasons (from our taxonomy) that a post is accurate via a checklist and 3) providing free-text rationales for why a headline is or is not accurate---on people's intention of sharing the headline on social media. 
From an experiment with 1668 participants, we find that both providing accuracy assessment and rationale reduce the sharing of false content. They also reduce the sharing of true content, but to a lesser degree that yields an overall decrease in the fraction of shared content that is false.

Our findings have implications for designing social media and news sharing platforms that draw from richer signals of content credibility contributed by users.
In addition, our validated taxonomy can be used by platforms and researchers as a way to gather rationales in an easier fashion than free-response.

\end{abstract}



\maketitle

\section{Introduction}

Social media has lowered the barrier to publishing content. While this empowerment of the individual has led to positive outcomes, it has also encouraged the fabrication of misinformation by malicious actors and its circulation by unwitting platform users \cite{vosoughi2018spread, grinberg2019fake}.
Given widespread concerns about misinformation on social media \cite{reed2018hate, whatsapp2019india, dixit2018whatsapp, bbcCovidMisinfo, conversationQAnon}, many researchers have investigated measures to counter misinformation on these platforms.
Some of these initiatives include detecting false or misleading information using machine learning algorithms \cite{castillo2011information, potthast2016clickbait, shu2017fake} and crowdsourcing~\cite{epstein2020will, pennycook2019fighting, allen2020scaling, kim2018leveraging, bhuiyan2020investigating}, identifying bad actors and helping good actors differentiate themselves\footnote{The Trust Project: https://thetrustproject.org/, Credibility Coalition: https://credibilitycoalition.org/} \cite{zhang2018structured}, and providing fact-checked information related to circulated news claims \cite{kriplean2014integrating, graves2016deciding, pennycook2020implied, yaqub2020effects}.
Social media companies themselves have also enlisted more contract moderators as well as third-party fact-checkers to remove or down-rank misinformation after publication~\cite{Facebook3rdpartyFactcheck}.

While credibility signals and fact-check flags provide useful filters and signals for readers, and algorithmic and human moderators help stop misinformation that is already spreading, none of these initiatives confront the underlying design of social media that leads to the proliferation of misinformation in the first place.
Part of the reason misinformation spreads so well is that social media prioritizes engagement over accuracy. For instance, a user who encounters an incorrect post and then leaves a comment refuting it may in fact be inadvertently disseminating it farther because the system considers the comment as engagement. A related driver is an emphasis on low barriers to sharing that allows users to share content without much attention to its accuracy or potential negative consequences, simply to receive social feedback or elevated engagement \cite{bazarova2015social, grinberg2017understanding}.

Thus, in this work, we begin to explore how one might alter social media platforms to better surface accuracy and evidence, not just engagement, when spreading content. To achieve this, we consider how to raise some small barriers to sharing in order to promote accurate content, without drastically changing the lightweight sharing practices typical in social media.
For example, sharing of any kind of content would likely plummet if platforms demanded that users receive substantial training in assessing information before being allowed to share, or that they perform extensive research on every post before sharing it. Instead, we look to interventions matched in scale to current sharing behavior such as clicking a ``like button'' or emoticon, adding a hashtag, or writing a comment. We therefore explore the potential impact of (i) requiring sharers to click a button to indicate whether they think a story is accurate or not when sharing it, (ii) requiring sharers to choose at least one tag from a small checklist indicating \emph{why} they consider it accurate, and (iii) writing a short comment explaining their accuracy assessment.
Introducing these interventions at posting time would encourage users to reflect on veracity before posting and perhaps reconsider posting an inaccurate piece of content. These assessments could also provide valuable information to other users seeking to form their own assessments of a post's accuracy.

In this work, we describe two studies completed using participants from Mechanical Turk motivated by these ideas. The first aims to develop a small set of categories that (i) cover the majority of reasons people consider content to be accurate or inaccurate and (ii) can be used accurately by them when providing assessments. This allows us to test as one of our interventions a checklist that can reasonably replace a free-text response.
The second study considers the impact of our three proposed interventions---accuracy assessment, reasoning ``categories'' provided by the first study, and textual reasoning---on sharing behavior.

\subsection{Research Questions}

Prior work has shown that priming people to think about accuracy when they are considering sharing a post can help reduce their likelihood of sharing falsehoods ~\cite{pennycook2020fighting, pennycook2021shifting}.
But little work has been done on how asking about accuracy could be integrated into platforms in a lightweight way.
In the first study, we examine how to capture people's rationale for a claim's (in)accuracy in a structured format. To create categories of rationales, we ask the following research question:
\begin{itemize}
  \item \textbf{RQ1:} What are the reasons why people believe or disbelieve news claims?
\end{itemize}

 In this study, we developed a taxonomy of reasons and presented a set of claims along with the taxonomy to participants to choose from, as they provided their rationales for the (in)accuracy of the news claims. We iterated on the taxonomy until participants were able to use it reliably to label their rationales.

Armed with our taxonomy of reasons, the second research question that we address is:
\begin{itemize}
  \item \textbf{RQ2}: How does asking people to provide accuracy assessments of news claims and their rationales affect their self-reported sharing intentions on social media?
\end{itemize}

Based on results from prior literature \cite{pennycook2021shifting, pennycook2020fighting}, our hypothesis is that asking people about content accuracy lowers their intention of sharing false content, but also of sharing true content to some degree. An ideal intervention would significantly reduce sharing of false content without much impacting the sharing of true. We further hypothesize that additionally asking people to provide their reasoning when evaluating the accuracy of content would help them even more with the discernment than simply asking for an accuracy judgement.

The study we conducted involved presenting news claims spanning different domains and partisan orientations to a set of participants and asking a subset if the claims were accurate and another subset their reasons for believing so in addition to accuracy assessment. For each news claim, participants indicated whether they would share it on social media.
We find that asking about accuracy decreases the likelihood of sharing both true and false news but of false news to a greater degree. We also find that additionally asking for rationale via free-text leads to a similar outcome, reducing sharing of both true and false headlines, but further decreasing the ratio of shared false headlines to true ones.
We delve deeper into providing rationales by comparing a condition where reasoning is provided in free-text to one where users must also select from a checklist of the reason categories taken from our taxonomy. We find that having people additionally work through the checklist of reasons does not further reduce ratio of shared false headlines to true ones compared to free-text reasoning alone. However, such structured rationales can be beneficial for integration into platforms as signals of credibility.

Our findings on the effects of the accuracy and reasoning nudges on sharing decisions point to potential designs for social media platforms to introduce at posting time to reduce sharing of misinformation. 
In addition, our taxonomy of rationales can be used to form a checklist for users to report their reasons for believing or disbelieving a claim in an easier fashion.
Because these inputs are \textit{structured annotations}, they provide a potential added benefit in that they can be \emph{aggregated} and presented to friends of the poster, for example indicating that ``2 friends think this post is true and 27 think it is false, 2 based on firsthand knowledge.'' 
This aggregate assessment could help warn a user about misinformation and could also help guide users to friends who are misinformed and could benefit from a conversation about the post.
We conclude by discussing these and other possible social media designs that could be introduced as a result of this work.

\section{Related Work}

We situate our study in prior work related to how misinformation spreads and what factors affect people's sharing behaviors, measures to combat misinformation, and factors influencing people's perceptions of content accuracy. 

\subsection{Spread of Misinformation}
Although misinformation is not a recent phenomenon \cite{posetti2018short}, the fairly recent use of online platforms for its dissemination has gained misinformation fresh attention. Researchers have focused on defining the problem space of misinformation \cite{wardle2017information, vraga2020defining} and determining how it has changed the societal and political arena \cite{bovet2019influence, shane2017fake}. A body of work examines data collected from online communities to study how people use social media to seek and share information, and how misinformation spreads through these communities \cite{oh2010exploration, del2016spreading}. For example, Starbird et al. investigate rumors that emerge during crisis events and report on the diffusion patterns of both false content and its corrections \cite{starbird2018engage, starbird2014rumors}. Vosoughi et al. analyze the spread of rumor cascades on Twitter and find that, among fact-checked articles, false content diffuses farther, faster, deeper, and more broadly than the truth \cite{vosoughi2018spread}. By analyzing tweets during and following the US 2016 election, Shao et al. find evidence that social bots played a role in spreading articles from low-credibility sources. In addition, they identify common strategies used by such bots, e.g., mentioning influential users \cite{shao2018spread}. Other work has examined echo chambers on social media and their selective exposure to information, e.g., in communities that relate to certain conspiracy theory narratives \cite{quattrociocchi2016echo, del2016spreading, schmidt2018polarization, mosleh2021shared}.

Another strand of research tries to understand how people's sharing behavior on social platforms is impacted by different aspects of the content or the platform. For instance, Vosoughi et al. report that people are more likely to share information that is novel, a characteristic that false content usually has \cite{vosoughi2018spread}. Pennycook et al. report that subtly priming people to be mindful of content accuracy before deciding whether to share the content helps lower their intention of sharing false information \cite{pennycook2021shifting, pennycook2020fighting}. They argue that this phenomenon happens because although people are generally good at discerning accuracy, when deciding whether to share content, they are less concerned with accuracy than other aspects of sharing such as the amount of social feedback they receive. Focusing their attention on accuracy can help alleviate the problem. This interpretation is consistent with prior work that reports people who rely more on their intuition and engage in less critical thinking are more susceptible to believing political fake news in survey experiments \cite{pennycook2019lazy} and in fact share news from lower quality sources on Twitter~\cite{mosleh2021cognitive}.

We extend this body of work by studying the effects of asking people to provide accuracy assessments as well as their reasoning for why a news story is or is not accurate on their intentions of sharing the content, quantifying the degree to which they reduce sharing of false (and true) content. If these nudges prove to be effective at curbing the sharing of inaccurate content, news sharing platforms and social media can benefit from implementing them.

\subsection{Measures to Counter Misinformation}

Another body of work has been investigating how to combat misinformation. For example, Bode and Vraga investigate the roles that authoritative and expert sources, peers of social media users, and platforms play in correcting misinformation~\cite{vraga2017using, bode2018see}. Other such studies investigate the impact of the wording of corrections~\cite{bode2018see, martel2021you} and explore identifying misinformation in its early stages using previously known rumors \cite{wu2017gleaning}, presenting linguistic and social network information about new social media accounts to help users differentiate between real and suspicious accounts \cite{karduni2019vulnerable}, and increasing media literacy \cite{haigh2019information}.

A related thread of work studies how social media users engage with fact-checking in the wild. Zubiaga et al. explore how rumors are diffused on social media and how users respond to them before and after the veracity of a rumor is resolved. They report that the types of tweets that are retweeted more are early tweets supporting a rumor that is still unverified. Once a rumor has been debunked however, users do not make the same effort to let others know about its veracity \cite{zubiaga2016analysing}. In another study on rumors, Shin et al. analyze rumors spread on Twitter during the 2012 election and find that they mostly continued to propagate even when information by professional fact-checking organizations had been made available \cite{shin2017political}.
Shin and Thorson analyze how Twitter users engage with fact-checking information and find that such information is more likely to be retweeted if it is advantageous to the user’s group (partisanship) \cite{shin2017partisan}. Other work has studied circumstances under which fact-checking is effective and find that social media users are more willing to accept corrections from friends than strangers \cite{margolin2018political, hannak2014get}. 
In addition, a Twitter field experiment found that being corrected by a stranger significantly reduced the quality of content users subsequently shared~\cite{mosleh2021perverse}.

Social media platforms have also been exploring ways to ameliorate their misinformation problem. Facebook for example, started showing red flags on certain posts to signal their lack of credibility. Such flags however, encouraged people to click on the content, causing Facebook to remove the flags in favor of adding links to related articles underneath the posts \cite{Quartz}. Facebook has also reported that it lowers the ranking of groups and pages that spread misinformation about vaccination \cite{facebookVaccine}. Platforms have recently turned to working with third-party fact-checkers to remove content that violate their community policies~\cite{Facebook3rdpartyFactcheck}. These measures in general force the platforms into a truth-arbitration role which is especially problematic in cases where policies have not had the foresight to predict all accounts of problematic posts or in grey areas~\cite{Facebooknapalm, Facebookkurdish}. We are interested in exploring ``friend-sourced'' methods in which the platforms are only responsible for delivering the assessments and not for making them.

We study the effects of two behavioral nudges, requesting accuracy assessments and rationales, on sharing false news as countermeasures that could be incorporated into social media platforms. To best leverage them, we also study how to capture people's rationales in structured form. We hypothesize that requesting users to assess accuracy of news headlines at sharing time acts as a barrier to posting, reducing sharing of false content but also of true content to a lesser degree. In addition, we hypothesize that further asking users to provide rationales for their accuracy assessments will result in a higher reduction in sharing of false headlines, and potentially of true headlines although to a lesser degree.

\subsection{Why People Believe News}
A body of work has been investigating the characteristics of posts or of people's interaction with them that affect their perceived accuracy. For instance, number of quoted sources \cite{sundar1998effect}, prior exposure to a claim \cite{pennycook2018prior}, official-looking logos and domains names \cite{wineburg2017lateral}, and post topic and author username \cite{morris2012tweeting} have been found to impact perceptions of news credibility, whereas the news publisher has surprisingly little effect on news accuracy ratings~\cite{dias2020emphasizing}. 
Pennycook et al. find that attaching warning flags to a subset of news stories increases the perceived credibility of those without flags \cite{pennycook2020implied}. By conducting interviews with participants whose newsfeeds were manipulated to contain false posts, Geeng et al. study why people do not investigate content credibility, e.g., because they have undergone political burnout \cite{geeng2020fake}.

Our study builds upon prior work by investigating self-reported reasons why people believe or disbelieve claims. We develop a taxonomy from these reasons and revise it iteratively until people untrained in the taxonomy are able to use it reliably to label their own rationales. We hypothesize that by leveraging these structured rationales and deliberating on all the dimensions of accuracy, users will be more discerning of false vs true content compared to if they provide unguided free-form reasoning. In addition, structured reasons have the added benefit that they could easily be integrated into platforms as signals of content credibility.

\section{Terminology and Methods}

In this section we introduce terminology that we will use to discuss and evaluate our interventions, some of which are also used in the course of the studies.

\subsection{Performance Metrics for Sharing Interventions}

Our overall goal is to study interventions at the moment of sharing content online. To evaluate these interventions, we seek a meaningful performance metric. Interventions that occur only when a user has decided to share can only \emph{prevent} some sharing, thus reducing the amount of sharing overall. An intervention that might be considered ideal would not prevent sharing of true content but would prevent all sharing of false content. More generally, it is useful to separately consider the degree to which an intervention reduces sharing of true content and the degree to which it reduces sharing of false. Previous work~\cite{pennycook2021shifting,pennycook2020fighting,pennycook2020implied} often assessed the performance of an intervention by comparing the change in the \emph{absolute difference} between the rate at which true and false content was shared. But here, we argue that an intervention which results in no change in the \emph{difference} in sharing rates can still be highly beneficial if it changes the \emph{ratio} of sharing rates. 

Consider for example a user who shared 20\% of the true content and 10\% of the false content they encountered. If the ``input stream'' of content they read were balanced between true and false, then they would share twice as much true content as false, meaning the ``output stream'' of content they shared would be 2/3 true to 1/3 false. Now suppose that an intervention decreases their sharing rate on both true and false content by 5\%, to 15\% and 5\% respectively. There is no change in the absolute difference in sharing rates, but the user's new output stream consists of 3/4 true content and 1/4 false. Going farther, if both rates are decreased by 10\% the output stream will contain only true content. 

Therefore, in assessing the effect of interventions, we focus on the (change in the) \emph{ratio} of sharing rates rather than the difference. If a user shares a fraction $f$ of false posts and a fraction $t$ of true posts, then an input stream with a ratio $r$ of false posts to true posts will yield an output stream with a ratio $fr/t$ of false to true. Thus, an intervention that reduces the \emph{discernment ratio} $f/t$ will improve the output false-true ratio regardless of the change in the \emph{difference} of sharing rates.
Of course this comes at a cost: a reduction in the overall sharing of true content. Different platforms may have different cost-benefit analyses of this trade-off. Note also that a portion of the benefit can be acquired at a portion of the cost by invoking the intervention on only a certain fraction of the shares.

On many social platforms, content one user consumes and shares is content that has been shared by others.  If each user's assessments improve the false-true ratio by a factor $f/t$, then over a chain of $k$ sharing steps the ratio is improved by a factor of $(f/t)^k$ overall; so even minor improvements accumulate exponentially.

\subsection{Veracity and Perceived Accuracy}

We now introduce relevant terminology.
Previous work has shown that personal deliberation can improve people's ability to distinguish accurate from inaccurate information~\cite{zhang2018structured}. And our interventions seek to engender different amounts of that deliberation. Initially, it is possible that users might choose to share certain information without even considering whether it is accurate. Our minimal intervention, simply asking a user whether a claim is accurate, already forces users to deliberate at least enough to answer the question. We expect that spending more time and effort in deliberation will generally improve a user's accuracy assessments. However, there is a limit to this improvement based on a user's available knowledge (and inaccurate knowledge) and rationality that will prevent their ever assessing perfectly. We use \emph{veracity} to refer to whether the claim is accurate or not independent of the user's knowledge. We use \emph{perceived accuracy} to refer to the user's initial answer
to whether they consider the claim accurate. We expect that subjective assessment of accuracy will correlate with objective accuracy, but imperfectly. Finally, we define the \emph{average perceived accuracy} of a claim to define the fraction of users who perceive the claim as accurate.

\section{Experimental and Study Design}

The objective of our first study (Taxonomy study) was to develop a taxonomy of self-reported reasons why people believe or disbelieve news claims. In the second (Nudge study), we investigated whether asking people to provide accuracy assessments and reasoning for the (in)accuracy of a news claim before they share it on social media nudges them to be more mindful of its credibility and if this nudge affects their sharing behavior. We further examined the effects of different instruments (a free-format text-box or a structured set of checkboxes) for capturing reasons on sharing news stories that are not credible. Our study was approved by our Institutional Review Board.

\subsection{Claims}

\begin{figure}[!t]
\centering
\begin{minipage}{.45\textwidth}
 \centering
 \includegraphics[width=\linewidth]{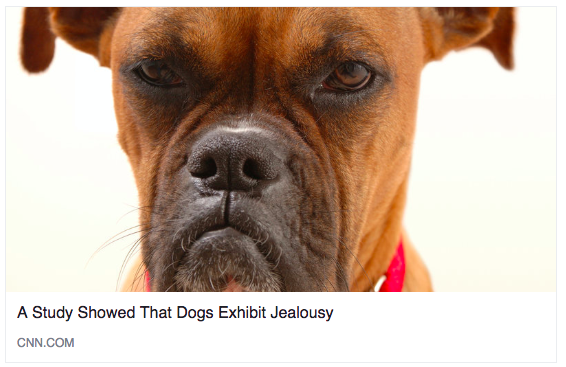}
 \captionof{figure}{An example of a headline in the study. Headlines were shown with an image, a lede sentence, and a source.}
 \label{fig:headlines}
\end{minipage}\qquad
\begin{minipage}{.45\textwidth}
 \centering
 \includegraphics[width=\linewidth]{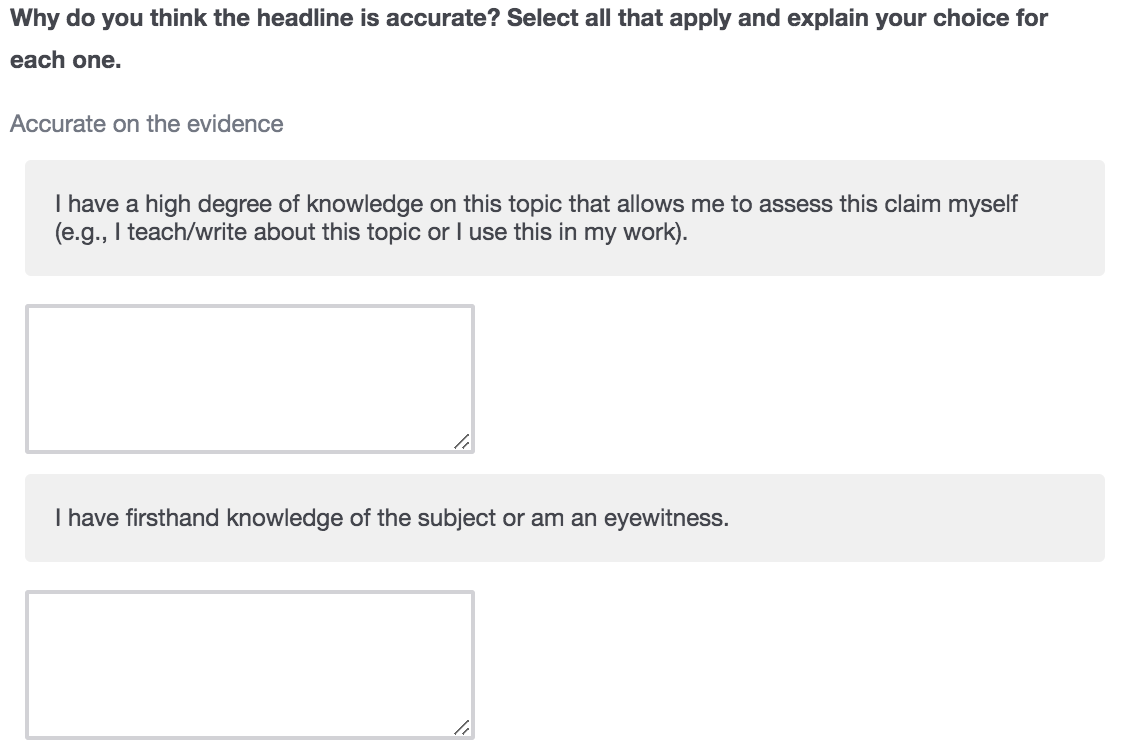}
 \captionof{figure}{The survey interface for the Nudge study iteration 4, where for each reason that a participant selected, they were required to fill in a text-box explaining their rationale. }
 \label{fig:checkbox_UI}
\end{minipage}
\end{figure}

We collected most of the claims we used in our studies from Snopes\footnote{https://snopes.com}, with a few from mainstream media. Each claim was presented with a picture, a lede sentence, and a source that had originally published an article on the claim, similar to news stories on social media (see Figure ~\ref{fig:headlines}) and also because users do not generally read entire articles and mainly limit their attention to headlines~\cite{manjoo2013you}.
The claims varied along different dimensions of veracity, domain, partisanship, and original source. For the claims that we collected from Snopes, we had the ground-truth that the website's fact-checkers had provided. We fact-checked the few that we collected from mainstream media by investigating the sources to which they referred. For domain, we chose claims that were either political or about science and technology, with claims in both domains covering a wide range of issues. Some of the claims were 
pro-Republican, some pro-Democratic, and others had no clear partisanship. The claims came from various sources including mainstream media, conspiracy websites, and social media. For the claims that had originally been circulated on social media such as Facebook, we displayed the source as ``Posted via Facebook.com''. 

Because we intended for our political claims to be relevant at the time of the study and not outdated, for each iteration of the study, we collected new political claims that had emerged or re-emerged within the past few months prior to the iteration. Selecting relevant headlines resulted in the iterations of the Taxonomy study having different but overlapping sets of claims, which supported our goal of a generalizable taxonomy.
Another set of claims was used for the Nudge study which was conducted in one iteration. These claims had a large overlap with those used in the last iteration of the Taxonomy study. We have provided this set in Appendix~\ref{section:headlines_table_nudge_study}.

In addition to relevance, we considered provenance when selecting claims for the study. We chose those claims for which Snopes had provided the originating source or the ones that it explicitly mentioned as being widespread rumors. For example, some claims had been requested to be fact-checked by Snope’s readership and therefore did not have a clear source or a place where they had emerged and therefore, we did not select these claims.
In addition, we filtered out those claims that were not factual (e.g., satire) because including them would have required presenting the item at the article and not the headline level.

\subsection{Participants}
We recruited U.S. based participants from Amazon Mechanical Turk. Across the 4 iterations of the Taxonomy study, 317 participants provided at least one (non-spam) answer to the news items presented to them. Of those, 305 completed the full survey and provided demographic information. The number of (non-spam) participants for the Nudge study was 1668, of whom 1502 provided demographic information. Of the 1807 participants across both studies who provided demographic information, 42\% were female. 47\% identified as Democratic, 25\% as Republican, and 26\% as Independent. They were distributed across a wide age range with a median of 35 years. The median income was \$40,000 to \$49,999.

The payment for the Taxonomy study was \$3. We determined the pay for the Nudge study (\$3.75) based on a soft-launch of the HIT with 100 workers which had a median completion time of 25 minutes. The soft-launch also revealed a high spam rate, leading us to limit the HIT to workers with a past approval HIT rating of higher than 95\% for all requesters.

\section{RQ1: Developing a Taxonomy of Reasons People Believe or Disbelieve Claims (Taxonomy study).}

We developed a taxonomy for people with no prior training to label their own rationales for why they (dis)believed a claim. We therefore, assigned a descriptive label to each category from a first person’s point of view (e.g., ``The claim is not consistent with my past experience and observations.''). A goal was for this description to be one that subjects could use correctly---that is, that participants would generally agree with each other, and with us, about the meaning of a particular category. We used multiple iterations of our study in order to achieve this goal, as described below.

\subsection{Procedure}

Through an online survey, participants were shown a series of 10 claims one at a time, with the claims randomly chosen from a pool. For each claim, the survey asked whether the claim was accurate or inaccurate, why, and how confident the participant was in their belief (4 point scale). Participants then answered another survey gauging their critical thinking~\cite{frederick2005cognitive}, statistical
numeracy and risk literacy~\cite{cokely2012measuring}, political knowledge, and attitudes towards science. These questions were drawn from research
studying users' judgements assessing claims~\cite{pennycook2019lazy}. Finally, participants answered demographics questions on political preference and theistic ideologies, among others.

We performed this study in 2 stages. To develop a preliminary taxonomy, we ran a first stage in which participants provided rationales for believing or disbelieving each of the claims via free-text responses. A total of 50 participants completed this stage of the study. We first divided the free-form responses that we collected into idea units, each being a coherent unit of thought \cite{strauss1987qualitative}. This resulted in 534 idea units. A member of the research team then conducted a first pass over the idea units and assigned preliminary categories to each using a grounded theory approach \cite{charmaz2007grounded}.

In the second stage of the study, we worked iteratively to refine our taxonomy, consolidating categories showing too much overlap and splitting some that represented distinct ideas. A particular goal was for the categories and their labels to align with how participants labeled their own responses. In this stage, for each claim, participants were asked to mark checkboxes corresponding to the reasoning categories they used and then to provide elaboration in text. To measure the alignment between our intended use for the categories and participants’ perception of them, a member of the research team with no knowledge of the users' checked categories assigned categories to the elaborated reasons. We then measured Cohen’s Kappa as a measure of the agreement between the categories selected by the research team coder and the ones participants had selected for their own responses. We conducted 3 rounds of this study, each time iterating over the categories.

The Kappa score in our initial iterations of the study was low which led us to revise the reason categories. However, we discovered that the low score was partly an artifact of the study design. In the initial iterations, participants were asked to first mark their reasons using the checkboxes and then to provide an aggregate explanation in one text-box. We noticed that the explanations did not cover all the selected reasons possibly because participants did not feel impelled to provide comprehensive explanations or that they deemed some checkboxes self-explanatory. We addressed this issue by modifying the survey interface so that for each checkbox that a participant selected, they were required to fill in a text-box explaining their reasoning (see Figure ~\ref{fig:checkbox_UI}).

Our attained agreement score in the 3rd iteration of this stage of the study was 0.63 which we measured across 729 idea units collected for 48 news claims. The score exceeded the recommended threshold for accepting the results \cite{landis1977measurement}.
Other scholars suggest a higher threshold for various tasks. However, while our attained agreement score may be lower than ideal, we deem it sufficient for this type of task.
The 4 iterations of the Taxonomy study spanned 13 months.

\subsection{Results}

Some categories emerging from the study would definitively determine that a claim was or was not accurate from the evaluator's perspective. We signalled the strength of these categories by grouping them under the name \textit{Accurate on the evidence} and \textit{Inaccurate by contrary knowledge}. Some rationales on the other hand were not conclusive but rendered a claim (im)plausible. For these, we used the terms \textit{Plausible} and \textit{Implausible} to indicate the strength. Other rationales were surmises and speculations. Although these rationales were not informative, we nonetheless wanted people to have the means to provide such rationales while indicating their lack of strength. We grouped these under the term \textit{Don't know}. 

We determined in the earlier iterations that the \textit{Don't know} categories were self-explanatory and the elaborations often repeated the terms. We therefore did not ask for elaborations when participants selected these categories in the final iteration.

Two categories emerged in the taxonomy that were used by participants in the initial stages as reasons a claim was inaccurate, but that we concluded were not reliable for deducing accuracy but rather belonged to other dimensions of credibility. One of these categories, \textit{The claim is misleading}, could be used for claims that for instance are accurate if taken literally, but are intentionally worded in such a way to lead the reader to make inaccurate deductions. 
Similarly, the category \textit{The claim is not from a trusted source} could be applicable to claims that are in fact accurate since sources of unknown reputation and even malicious ones publish accurate information mixed with false content or propaganda~\cite{starbird2018ecosystem}.
Therefore, we separated these two categories from the rest and requested that participants evaluate each claim on these two signals regardless of whether they judged the claim as accurate. Tables~\ref{tab:accurate_categories}, ~\ref{tab:inaccurate_categories}, and ~\ref{tab:other_signals} show the full taxonomy.

\subsubsection{Explanations of Certain Taxonomy Categories.}
Here we elaborate on some of the categories that were not self-explanatory.
Throughout the paper, where we present participants' free-text responses, we identify them with a string of the form ``p-'' + a participant number to preserve their anonymity. If a participant is from an iteration other than the final, we have concatenated the following string to the end of their identifier: ``-'' + iteration number.

\paragraph{The Claim Is (Not) Consistent with My Past Experiences and Observations.}
One of the most cited rationales for a claim’s (in)accuracy, was that it was (not) consistent with the participant’s past experience and observations. This assessment at times resulted from the participant’s general knowledge of how an authority, e.g., the law, operates: \textit{``I am pretty sure that to sign up for these benefits you would need a social security number.'' (p-9)} ---\textsf{Claim: Seniors on Social Security Have to Pay for Medicare While `Illegal Immigrants' Get It Free}.

At other times, the rationale was based on whether the assertion in the claim matched the subject of the claim’s past profile or pattern of behavior: \textit{“Joe Biden has a history of gaffes that come off this silly.” (p-79)} ---\textsf{Claim: Joe Biden Said: ‘Poor Kids Are Just as Talented as White Kids’}.

Sometimes the assessment referred to whether the claim confirmed or contradicted the customary state of world as the participant perceived it: \textit{“It has been my experience that attitudes on sexual orientation have changed considerably” (p-53)} ---\textsf{Claim: Age Matters More than Sexual Orientation to U.S. Presidential Voters, Poll Finds}.
This rationale also emerged in cases where the participant had heard about similar claims before and although the claim’s accuracy had not been established, the repeated encounter made it seem plausible:
\textit{“I've been hearing this statement since I was a child, snowflakes are like fingerprints. There are none that are identical.” (p-32-3)}---\textsf{Claim: No Two Snowflakes Are Exactly Alike}.

This phenomenon has also been reported in \cite{pennycook2018prior}, where Pennycook et al. found that even a single prior exposure to a headline increases subsequent perceptions of its accuracy. Surprisingly, the illusory truth effect of repeated false statements influences people across the political spectrum, i.e., even if their ideological beliefs disagree with the statements \cite{murray2020ve}.

In fact, in the earlier iterations of the taxonomy, \textit{Familiarity of Claim} was a separate category from \textit{Consistency with Past Experiences and Observations}. However, we merged the two because in many instances participants could not make the distinction.

\begin{footnotesize}
\begin{table}[t]
\caption{Taxonomy of reasons why people believe news claims.}
\label{tab:accurate_categories}
\begin{minipage}{\columnwidth}
\begin{center}
\footnotesize
\begin{tabular}{|c|p{3.3cm}|p{9.2cm}|}
 \hline
 \rowcolor{gray!28} & \textbf{Category} & \textbf{Example}\\
 \hline
 
 \multirow{5}{*}{\rotatebox[origin=c]{90}{\parbox[c]{8cm}{\centering Accurate on the evidence}}} & 
 I have a high degree of knowledge on this topic that allows me to assess this claim myself (e.g., I teach/write about this topic or I use this in my work). (N=2) & 
 \textit{``Global warming is really happening around us and we must stop it. I have researched it for some time now.'' (p-3)} \newline
 \textsf{Claim: Global Sea Ice is at a Record Breaking Low.}
 \\

 \cline{2-3}
 & I have firsthand knowledge of the subject or am an eyewitness. (N=23) &
 \textit{``My own dog does this when other dogs come into the picture, I can see her getting jealous for my attention.'' (p-60)} \newline
 \textsf{Claim: A Study Showed that Dogs Exhibit Jealousy.}
 \\
 
 \cline{2-3}
 & My other trusted sources (besides the source of this article) confirm the entire claim. (N=54) & 
 \textit{``I've read numerous articles from Huffington Post, Buzzfeed, and Mashable about this happening.'' (p-26)} 
 \newline
 \textsf{Claim: Some Phone Cameras Inadvertently Opened While Users Scrolled Facebook App.}
 \\
 
 \cline{2-3}
 & The claim is from a source I trust. (N=49)& 
 \textit{``I do trust the Washington Post to report accurately.'' (p-47)} \newline
 \textsf{Claim: Gun Violence Killed More People in U.S. in 9 Weeks than U.S. Combatants Died in D-Day} [source: WashingtonPost.com].
 \\
 
 \cline{2-3}
 & Evidence presented in the article corroborates the claim. (N=9) & 
 \textit{``The mountain peaks in the background look the same. I think the claim is very likely to be true.'' (p-23)} \newline
 \textsf{Claim: These Photographs Show the Same Spot in Arctic 100 Years Apart.} [The claim is presented with two juxtaposed photos, one showing mountains covered by glaciers, and in the other the glaciers have almost completely melted.]
 \\

 \cline{1-3}
 \multirow{1}{*}{\rotatebox[origin=c]{90}{\parbox[c]{1.5cm}{\centering Plausible}}} & 
 The claim is consistent with my past experience and observations. (N=120) & \textit{``This seems fairly consistent. The media seems to only report when Trump does wrong, even from the start.'' (p-71)} \newline
 \textsf{Claim: President Trump's Awarding of a Purple Heart to a Wounded Vet Went Unreported by News Media.}\\
 
 \cline{1-3}
 \multirow{1}{*}{\rotatebox[origin=c]{90}{\parbox[c]{3cm}{\centering Don't know}}} & 
 I’m not sure, but I want the claim to be true. (N=32) &
 \textit{``It is an interesting claim so I would hope that it was true but I've never heard of YUP and Twitter isn't very reliable.'' (p-46-3)}\newline
 \textsf{Claim: There is a Point in the Ocean Where the Closest Human Could Be an Astronaut.} [The picture presented with the claim shows a tweet explaining the claim from the source YUP.]
 \\
 \cline{2-3}
 
 & I was just guessing. (N=104) & \textit{``I have no knowledge of the headlines, but it seems plausible that it could be true based off just a guess.'' (p-38-3)}\newline
 \textsf{Claim: There Are More Trees on Earth Than Stars in the Milky Way.}
 \\
 \cline{1-3}
 & Other (N=14) & \textit{``It's probably not the whole story, and is probably connected to the lack of federal recognition of same sex marriages. As in because the marriages aren't recognized, the adoption of foreign national children by those couple as a couple aren't [sic] recognized, so the child can't be naturalized, etc.'' (p-37)}\newline
 \textsf{Claim: The Trump Administration Is Denying U.S. Citizenship to Children of Same-Sex Couples.}
 \\
 \hline
 \end{tabular}
 \end{center}
 \end{minipage}
 \end{table}
 \end{footnotesize}

\begin{footnotesize}
\begin{table}[t]
\caption{Taxonomy of reasons why people disbelieve news claims.}
\label{tab:inaccurate_categories}
\begin{minipage}{\columnwidth}
\begin{center}
\footnotesize
\begin{tabular}{|c|p{3.3cm}|p{9.2cm}|}
 \hline
 \rowcolor{gray!28} & \textbf{Category} & \textbf{Example}\\
 \hline
 
  \multirow{5}{*}{\rotatebox[origin=c]{90}{\parbox[c]{4.5cm}{\centering Inaccurate by contrary knowledge}}} & 
 I have a high degree of knowledge on this topic that allows me to assess this claim myself (e.g., I teach/write about this topic or I use this in my work). (N=3) & 
 \textit{``Asylum seekers can only apply on US soil. I've worked with immigrants abroad for an extended period of time.'' (p-2)} \newline
 \textsf{Claim: Asylum-Seekers Can Apply at U.S. Embassies Abroad.}\\
 
 \cline{2-3}
 & I have firsthand knowledge on the subject or am an eyewitness. (N=7) &
 \textit{``I watched the news coverage.'' (p-67)}
 \newline
 \textsf{Claim: ABC, CBS, and NBC Blacked Out Pam Bondi's Legal Defense of Trump during His Impeachment Trial.} \\
 
 \cline{2-3}
 & The claim contradicts some information related to the case that I know from trusted sources. (N=49) &
 \textit{``I think that number is almost equal to the total number of homicides in the US which is a ridiculous notion.'' (p-80)}\newline
 \textsf{Claim: 10,150 Americans Were Killed by Illegal Immigrants in 2018.}\\
 
 \cline{1-3}
 \multirow{5}{*}{\rotatebox[origin=c]{90}{\parbox[c]{7cm}{\centering Implausible}}} & 
 The claim is not consistent with my past experience and observations. (N=46) & 
 \textit{``The man is a showman, there's no way he'd do something like this without letting anyone know about it.'' (p-30)} \newline
 \textsf{Claim: President Trump's Awarding of a Purple Heart to a Wounded Vet Went Unreported by News Media.}\\
 
 \cline{2-3}
 & If this were true, I would have heard about it. (N=91) & 
 \textit{``I feel [something] like this would have been a huge story that would have been carried on many national news networks.'' (p-39)} \newline
 \textsf{Claim: US Intelligence Eliminated a Requirement That Whistleblowers Provide Firsthand Knowledge.}
 \\
 
 \cline{2-3}
 & The claim appears to be inaccurate based on its presentation (its language, flawed logic, etc.). (N=14) & 
 \textit{``This tweet follows the standard "ah! everybody panic!" format you see for unsubstantiated information. Also, there is no link to a source.'' (p-79)} \newline
 \textsf{Claim: Presidential Alerts Give the Government Total Access to Your Phone.} [The accompanying picture shows a tweet by John McAfee warning about the issue.]
 \\
 \cline{2-3}
 & The claim appears motivated or biased. (N=90) & 
 \textit{``Not saying which library or for what reason leads people to come up with their own conclusions.'' (p-30)} \newline
 \textsf{Claim: Biden's Campaign Demanded an American Flag Be Removed from a Library.}
 \\
 \cline{2-3}
 & The claim references something that is impossible to prove. (N=12) & 
 \textit{```Most sane of all' is a hard metric to measure.'' (p-29)} \newline
 \textsf{Claim: A Scientific Study Proved that ``Conspiracists'' Are ``The Most Sane of All''.}
 \\
 
 \cline{1-3}
 \multirow{1}{*}{\rotatebox[origin=c]{90}{\parbox[c]{2.5cm}{\centering Don't know}}} & 
 I’m not sure, but I do not want the claim to be true. (N=99) &
 \textit{``Knowing that I don't want to believe but aware of Biden's inaccurate pronouncements, I just chose not to give this any credence because, it is irrelevant.'' (p-37-3)}\newline
 \textsf{Claim: Joe Biden said the Mass Shootings in El Paso and Dayton Happened in `Houston' and `Michigan'.}
 \\
 \cline{2-3}
 
 & I was just guessing. (N=74) & \textit{``I am purely guessing here because I don't know. I'd rather not believe something that is true than believe something that is actually false.'' (p-36-3)}\newline
 \textsf{Claim: Mueller Concluded Trump Committed `No Obstruction' in the 2016 Election Probe.}
 \\
 \cline{1-3}
 & Other (N=61) & \textit{``If migrants could leave at any time, why wouldn't they leave. What would be the point in detaining them and sending them to a center if they were free to do as they please in the first place?'' (p-9)}\newline
 \textsf{Claim: Migrants `Are Free to Leave Detention Centers Any Time'.}\\

 \hline
 \end{tabular}
 \end{center}
 \end{minipage}
 \end{table}
 \end{footnotesize}

\begin{table}[h]
\caption{Other Signals of credibility that do not necessarily render a claim accurate or inaccurate.}
\label{tab:other_signals}
\begin{minipage}{\columnwidth}
\begin{center}
\footnotesize
\begin{tabular}{|p{2.8cm}| p{10.2cm}|}
 \hline
 \rowcolor{gray!28} \textbf{Category} & \textbf{Example}\\
 \hline
 The claim is misleading. ($N_\textit{Accurate}=23$, $N_\textit{Inaccurate}=57$)  & \textit{``It might be that women end up with DNA-containing fluid/skin cells during sex, but not permanently. Or during pregnancy some of DNA from the fetus (which would contain some of the male partner DNA) might end up in the woman's blood. But not every partner's.'' (p-37)}\newline
 \textsf{Claim: Women Retain DNA From Every Man They Have Ever Slept With.}
 \\
 \cline{1-2}
 The claim is not from a source I trust. (N=4) & \textit{``NBC has a habit of exaggerating to obtain ratings.'' (p-88)}\newline
 \textsf{Claim: Watchdog: ICE Doesn't Know How Many Veterans It Has Deported} [source: NBCNews.com].
 \\
 \hline
 \end{tabular}
 \end{center}
 \end{minipage}
 \end{table}

\paragraph{The Claim Appears to Be Inaccurate Based on Presentation.}

Participants reported how a claim looks as a factor impacting their perception of the claim’s accuracy. They referred to sensational language---\textit{“It's too much like a sensationalist tabloid headline.” (p-10)}, grammatical errors---\textit{“I really have no idea if this is true or not but the language seems weird.” (p-46)}, clickbait titles---\textit{“The title seems to be clickbait” (p-27)}, and quality of presented artifacts---\textit{“The image looks like a generic image that is commonly used with clickbait. The website referenced is "cityscrollz.com" which has a stylized spelling and does not seem to be a reliable news source.” (p-65)} as indicators of an article’s falsehood. Interestingly, some of these factors are among the set of indicators for evaluating content credibility suggested by Zhang et al. \cite{zhang2018structured}. While Zhang et al.’s proposed indicators are intended for evaluating both accurate and inaccurate content, in our study, this type of rationale was cited only as an argument for refuting a claim. One explanation is that because we only showed headlines and not full articles, participants may have interpreted the presence of these factors as red flags invalidating the claim, but their absence simply calling for more investigation.

\paragraph{The Claim Appears Motivated or Biased.}
Sometimes participants determined that a claim was false because it seemed to advance a particular agenda. In most cases, they did not know about the accuracy of the particular claim, but based their assessment on their prior familiarity with the general topic or the source of the claim:
\textit{“Fox news is biased and probably doesn't like whatever the "new way forward" is and wants to focus on the negatives.” (p-45)} ---\textsf{Claim: The New Way Forward Act Would Protect Criminals from Deportation} [source: FoxNews.com].
This type of rationale generally surfaced for partisan issues and claims that were associated with particular groups and movements: 
\textit{"The source has a liberal bias, and likely is including suicides as `violence' which most will interpret as person on person violence." (p-24)} ---\textsf{Claim: Gun Violence Killed More People in U.S. in 9 Weeks Than U.S. Combatants Died in D-Day} [source: WashingtonPost.com].

\subsubsection{Common Misalignments and Transformation of the Categories.}

The taxonomy categories underwent multiple iterations of transformation based on the misalignment between our expected use of the categories and how participants used them. Here we elaborate on some examples of this misalignment as potential areas for future refinement.

\paragraph{My Trusted Sources Confirm the Entire Claim.} This category was intended for cases where a participant had heard about the claim from other sources or had learned about it from an authority (e.g., at school). Sometimes, participants believed that they had heard about the claim before but that they did not fully remember the particulars of the claim they had heard or from whom they heard it, but that having encountered it nonetheless made them more accepting of its plausibility. In these cases, they often used the category \textit{The claim is consistent with my experience and observations}, as seen in the following example: \textit{``It seems like something I have read/heard before in the past.'' (p-16-3)} ---\textsf{Claim: U.S Sen. Lindsey Graham Once Said a Crime Isn't Required for Impeachment.}
Therefore, it appears that the degree of confidence in one's rationale in addition to the type of the rationale can shift the assigned label from one of these categories to the other.

\paragraph{I Have a High Degree of Knowledge on this Topic that Allows Me To Assess the Claim Myself.} In the earlier iterations of the taxonomy, this category was referred to as \textit{I Have Specific Expertise on the Subject}. However, we discovered that the definition of expertise varied across participants, as demonstrated by this example that was labeled by the participant as belonging to the expertise category:
\textit{``I have seen this nearly exact headline on social media, and was curious about its claims. Turns out that, from what I read on the government site, that this headline is misleading.'' (p-28-3)}
In the subsequent iterations, in addition to refining the name of the category, we added the example \textit{I teach/write about this topic or I use this in my work} to better convey the intended bar for expertise.

\paragraph{I Have Firsthand Knowledge on the Subject or Am an Eyewitness.} Participants occasionally said they had investigated the sources of a claim or had heard another source verify or refute the claim and thus had firsthand knowledge: \textit{``I saw it from reliable sources, I witnessed the news articles myself firsthand.'' (p-67)}
However, our intended category for such occasions would be \textit{My other sources confirm the entire claim} if considered accurate, and \textit{The claim contradicts some information related to the case that I know from trusted sources} if considered inaccurate.

\subsubsection{Demographics.} We investigated whether the demographics of participants have an effect on the types of rationales that they produce. We found that the distribution of rationales differ statistically across genders and present details in the Appendix Section~\ref{section:demographics_taxonomy_study}.

\subsection{Discussion of Taxonomy}

Some categories in Table~\ref{tab:other_signals} may appear similar to other rationales that can render a claim implausible and need to be further distinguished. One such pair is a claim being misleading and its appearing to be inaccurate based on its presentation (e.g., use of sensational language). In the absence of other information, sensational language renders the claim implausible, i.e., fails to convince the user that the claim is accurate. The claim being misleading however, is not a reason why the claim should be accurate or inaccurate, but exists in addition to the accuracy dimension. Users could consult other rationales to determine the accuracy or plausibility of the claim, and determine that for instance, although the claim is accurate, the accurate information pieces are chosen and the article crafted in such a way as to imply a particular inaccurate message. Another such pair is the claim appearing motivated or biased and its source being one that the user does not trust. To determine whether a claim is motivated or biased, users often resort to such information as their familiarity with the subject matter or their prior knowledge of the agenda of the source as well as their inferred bias of the claim to determine whether the truth has been twisted in the claim. Therefore, when the message of the claim agrees with the bias of the source, users see it as an indication that the claim may in fact not be accurate. A separate dimension of credibility is whether the source is not trusted by the user. For instance, a user who has stated they do not trust Fox News may in fact assess Fox’s claim that Biden won the presidential election in Arizona~\cite{fox2020BidenArizona} as accurate, the claim as not biased, while maintaining that the claim is from a source they do not trust.

We now discuss some general issues that arose in this study.

There is an important logical asymmetry between being consistent and inconsistent with past experience. Inconsistency with the past does offer some level of indication that the claim is inaccurate. However, given the tremendous number of things that \emph{could} happen consistent with the past, consistency offers little evidence that a claim is accurate---instead, it only \emph{fails} to provide evidence that the claim is not accurate. In general, many participants seemed not to make this distinction, using consistency with the past as sufficient evidence that a claim is true. Because participants used this category often, system designers may feel compelled to make it available. But those designers might want to consider treating this category as indicating that the user \emph{does not know} whether the claim is accurate, rather than indicating accuracy.

The confusion of lack of refutation with accuracy offers opportunities for manipulation. It suggests that a subject might tend to believe that a politician voted yes, and equally that a politician voted no, simply because they saw one or the other headline, without any other evidence. 

In a similar vein, some subjects treated \textit{The claim is not from a source I trust} as a reason to consider a claim false. Related work shows that alternative media sources borrow content from other sources including mainstream media~\cite{starbird2018ecosystem}. Therefore, it is important to bring users to this realization that sources of unknown or low reputation may in fact publish accurate content.
While we observed that users can rather reliably use the taxonomy in its current format to characterize their rationales, the taxonomy can still benefit from further refinement, which we leave to future work.

\section{RQ2: Effects of Providing Accuracy Reasons on Sharing Behavior (Nudge study).}

We hypothesized that asking people to reflect on the accuracy of a news story before they share it on social media would help prevent sharing news stories that are not credible. 
We additionally hypothesized that getting people to consider their rationales would help with their deliberation. 
For these purposes, we used the taxonomy that we developed in the Taxonomy study as one option to nudge people to consider possible rationales, along with a second free-text option.

\subsection{Method}
\label{section:nudge_method}

\begin{figure}[!t]
\centering
\begin{minipage}{.45\textwidth}
 \centering
 \includegraphics[width=\linewidth]{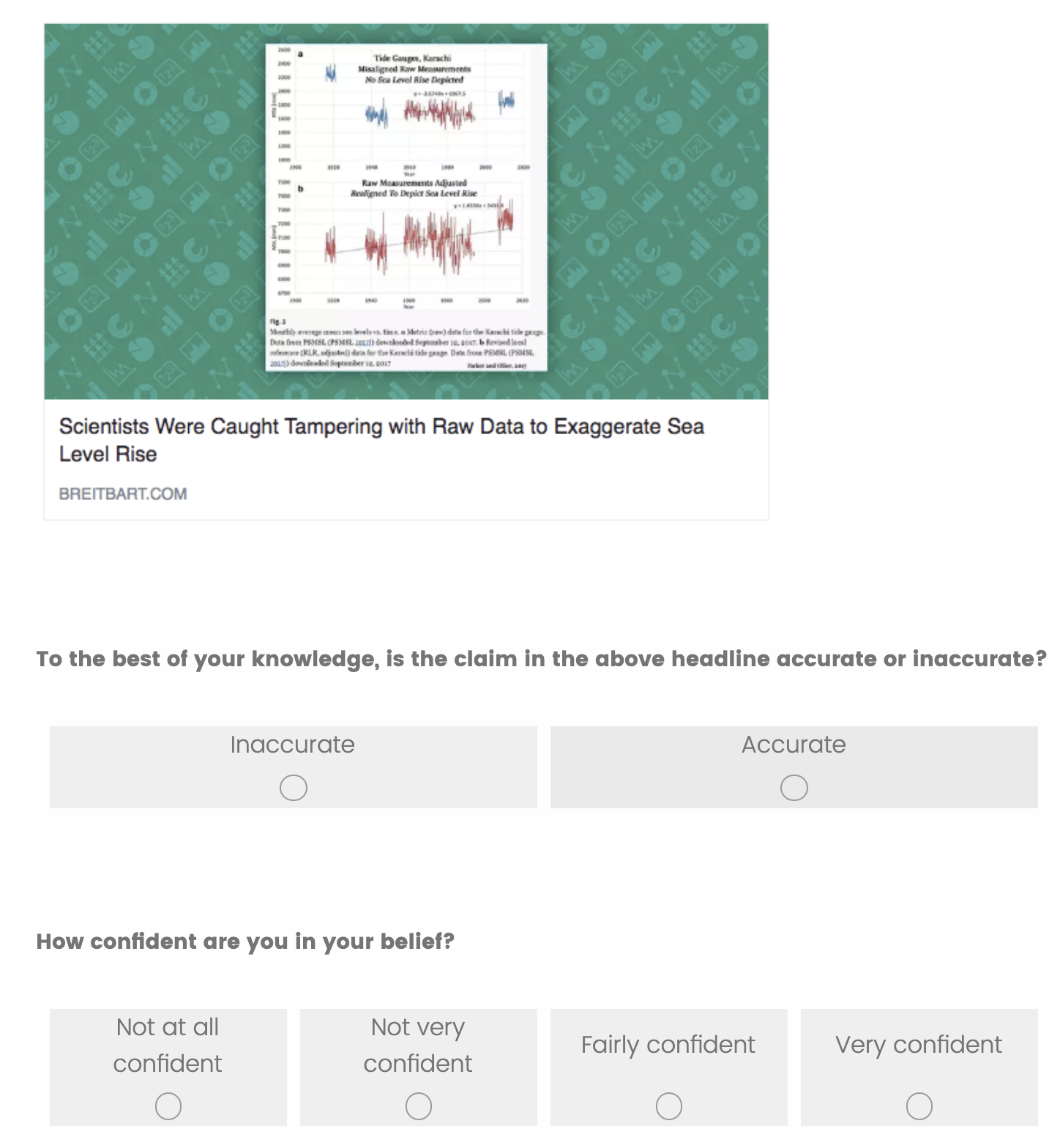}
 \captionof{figure}{The UI for how accuracy and confidence questions were presented to a participant along with an example of a headline.}
 \label{fig:accuracy_UI}
\end{minipage}\qquad
\begin{minipage}{.45\textwidth}
 \centering
 \includegraphics[width=\linewidth]{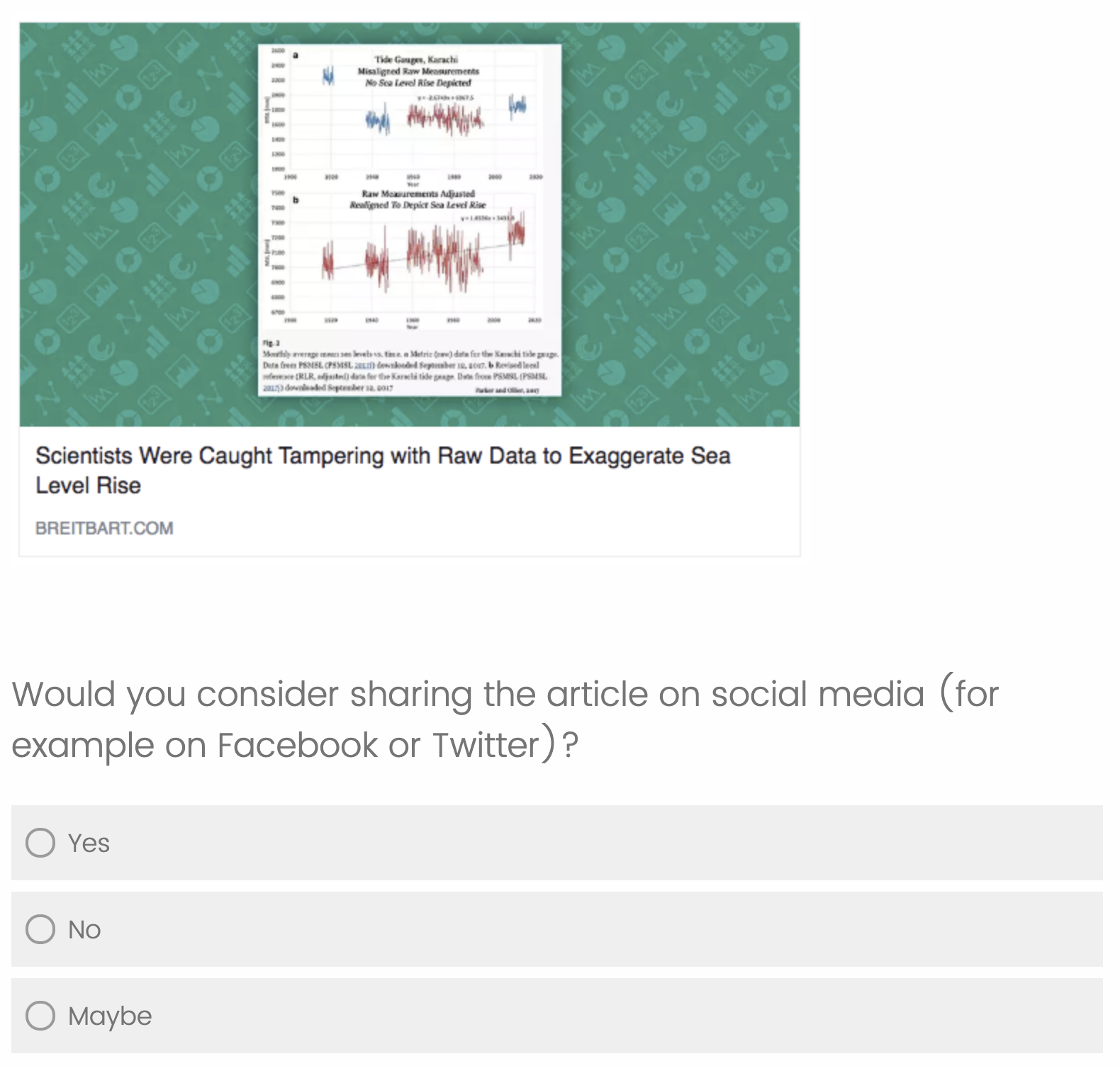}
 \captionof{figure}{The UI for how we asked users whether they would consider sharing the headline presented to them.}
 \label{fig:share_UI}
\end{minipage}
\end{figure}

Table~\ref{tab:conditions} summarizes our experimental conditions.
Similar to the Taxonomy study, participants were shown a series of 10 claims one at a time via an online survey. Headlines were randomly drawn from a set of 54 headlines. Of the pool of headlines, 24 were true, 25 false, and 5 were assessed as being a mixture of true and false. 
If a participant was in any of the treatment conditions, for each claim, the survey would ask whether the claim was accurate or inaccurate and how confident the participant was in their belief (4 point scale),
displayed in Figure~\ref{fig:accuracy_UI}.
If the participant was in one of the reasoning conditions, the survey would additionally ask why they believed the claim was (in)accurate. At the end of each item, all participants were asked if they would consider sharing the article on social media, with options Yes, Maybe, and No,
displayed in Figure~\ref{fig:share_UI}.
We also followed up with another question asking why they would (not) consider sharing the article. 
Following the claims, participants answered a survey asking how they would verify the accuracy of a headline like what they saw and how comfortable they were with asserting their judgments publicly.
Then they answered partisanship questions for each claim that they had previously seen: ``Assuming the above headline is entirely accurate, how favorable would it be to Democrats versus Republicans?'' (5 point scale). 
The survey contained similar post-task questions as the Taxonomy study.
The full questionnaire is included in the Supplementary Materials.
The Nudge study was completed in 12 days.

\begin{table*}[!t]
\setlength{\aboverulesep}{0pt}
\setlength{\belowrulesep}{0pt}
\setlength{\extrarowheight}{.35ex}
\newcolumntype{g}{>{\columncolor{gray!16}}c}
\setlength{\tabcolsep}{6pt}
\caption{Experimental conditions of the Nudge study. Table shows participants in which conditions were presented with questions to assess the accuracy of claims, provide their reasoning for why the claim is (not) accurate, and whether we used the taxonomy we developed in the Nudge study to capture their reasoning.}
\label{tab:conditions}
\centering
\begin{tabular}{gccl}
 \toprule
 \rowcolor{gray!16} Cond. & {Assessed accuracy} & {Provided reasoning} & {Reasoning format} \\
 \midrule
 1 & -- & -- & -- \\
 2 & \ding{51} & -- & --\\
 3 & \ding{51} & \ding{51} & Free-form text\\
 4 & \ding{51} & \ding{51} & Checkbox of taxonomy categories + text\\
 \bottomrule
\end{tabular}
\end{table*}%

\subsection{Results}
\label{section:nudge_main_results}

Our dataset from the Nudge study contained 21,113 datapoints of which we identified 5,403 as spams by investigating their associated free-text responses. The responses that we labeled as spam were copy pastes, sometimes with minor modifications, of the headline title or unrelated answers to the question (e.g., responding ``good'' to all the questions). Other responses that we labeled as spams had glaring grammatical errors and were mismatched between the participant's accuracy assessment and their text. In these cases, we deemed that the participant had not received the intended treatment and therefore, we treated their response as a spam.
Exclusions were applied at the participant level as whenever we determined that a datapoint was suspicious of being spam, we examined all the other datapoints submitted by the same participant as well as their responses to the survey that followed the claims, described in Section~\ref{section:nudge_method}. In almost all of the cases, the qualities that disqualified a datapoint were present in all of that participant’s responses, and therefore, we labeled all the participant’s submitted datapoints as spam. 		

A datapoint that we discarded as spam for example was the following: \textit{``the claim will be dangerous.''}---in response to \textit{Why do you think the claim is inaccurate?}; \textit{``it gives the fear about the treatment.''}---in response to \textit{Why would you not consider sharing it?}; both from the same datapoint; \textsf{Claim: Tibetan Monks Can Raise Body Temperature With Their Minds.}
We also excluded 6 datapoints where participants had technical issues.
The datapoints that we included in the analyses were collected from 1,668 participants.
Participants did not always complete all 10 claims due to dropout or technical issues; also, we remove from our analyses some datapoints participants labeled for headlines whose ground truth was neither completely true nor false (mixture), which were collected for exploratory purposes.
From the datapoints that we included in the analyses, 3,740 were in condition 1, 3,977 in condition 2, 3,405 in condition 3, and 3,118 in condition 4.

\subsubsection{Models.}

To test the effect of the nudges and their interaction with objective or subjective veracity of headlines, we fit two types of models to our dataset, both with share intention as the dependent variable. One was a linear mixed effect model for which we assigned values of 0, 0.5, and 1 to the share decisions of ``No'', ``Maybe'', and ``Yes'' respectively. The other was a cumulative link mixed model which treated the share decisions as ordinal. Results were consistent between the two models. Because the linear model is more straightforward to interpret, we discuss the results of this model below and leave the results of the cumulative link model to the Appendix section~\ref{section:cumulative_link_model}.

\paragraph{Veracity Model.}
\label{section:veracity_model}
To test the effect of nudges and their interaction with objective veracity of headlines, we developed a linear mixed effect model with sharing intention as the dependent variable and our study treatments as independent variables. The treatments were whether participants were asked about accuracy, whether they were asked about reasoning, and whether the reason checklist was presented to them. The model also included the veracity of the headline and the interaction between veracity and each of the treatments as independent variables. We fit this model to the whole dataset. We included participant identifier and the headline the participant had assessed as random effects in our models. The inclusion of these random effects accounts for the non-independence between the data points provided by one participant or for one headline and captures the variance in the intercepts between participants or headlines. We used the function ``lmer'' from the R package "lme4" to define the model. 
We refer to this model as the \textbf{veracity model}:
\begin{multline}
    \text{share} \sim \text{veracity} \times (\text{accuracy condition} + \text{reasoning condition} +\text{reasoning format)} \\ + (1|\text{participant}) + (1|\text{claim})
\end{multline}
We also developed another more refined veracity model with the demographics of participants as control variables, discussed in Appendix~\ref{section:nudge_demographics}. The effects we observed for headline veracity as well as our treatments in the model discussed in the Appendix were consistent with the results we observed for the model outlined in this section.

\paragraph{Perceived Accuracy Model.}

Although the ultimate desired sharing behavior is for people to share content that is in fact true and refrain from sharing objectively false stories, the best achievable outcome for behavioral nudges that encourage deliberation is to guide people to come to a better discernment \emph{based on what they already know}. One realization for example, could be that maybe after all, they do not know what they previously had taken for granted. Therefore, in addition to examining how the nudges affect sharing of objectively true and false content, we investigate how they interact with headlines that participants had initially believed to be true or false, indicated by their accuracy assessments.

Therefore, to test how the treatments and their interaction with a participant's initial accuracy assessment affect sharing intentions, we fit a model similar to the one described in~\ref{section:veracity_model} but included perceived accuracy, i.e., participant's assessment of the accuracy of the headline, rather than veracity as the independent variable. Because we did not have accuracy assessments from participants in the control (condition 1), we fit this model to the data from conditions 2, 3, and 4. The treatments that we included as independent variables were whether participants were asking about reasoning and whether we presented the reason checklist to them. We refer to this model as the \textbf{perceived accuracy model}:
\begin{multline}
    \text{share} \sim \text{perceived accuracy} \times (\text{reasoning condition} +\text{reasoning format)} + \\ (1|\text{participant}) + (1|\text{claim})
\end{multline}
\subsubsection{Findings.}

We performed a Wald Chi-Square test on each of the fitted models to determine if our explanatory variables were significant. The tests revealed that the effect of veracity in the veracity model and the effect of perceived accuracy in the perceived accuracy model were both significant [$\chi^2(1)=34.03$, $p<0.001$ for veracity, $\chi^2(1)=2025.12$, $p<0.001$ for perceived accuracy]. Post-hoc Estimated Marginal Means tests revealed that participants were more likely to share an objectively true rather than a false headline [$z=4.98$, $p<0.001$]. Similarly, they were more likely to have the intention to share a headline that they assessed as accurate [$z=37.70$, $p<0.001$].
We report the result of the tests for each of our study interventions in sections that follow.

Throughout the paper, we present figures showing the means of our outcome measures across conditions. The error bars in these figures are standard errors around the mean.

\paragraph{Effect of Providing Accuracy Assessments.}

\begin{figure}[t]
 \centering
 \includegraphics[width=0.8\linewidth]{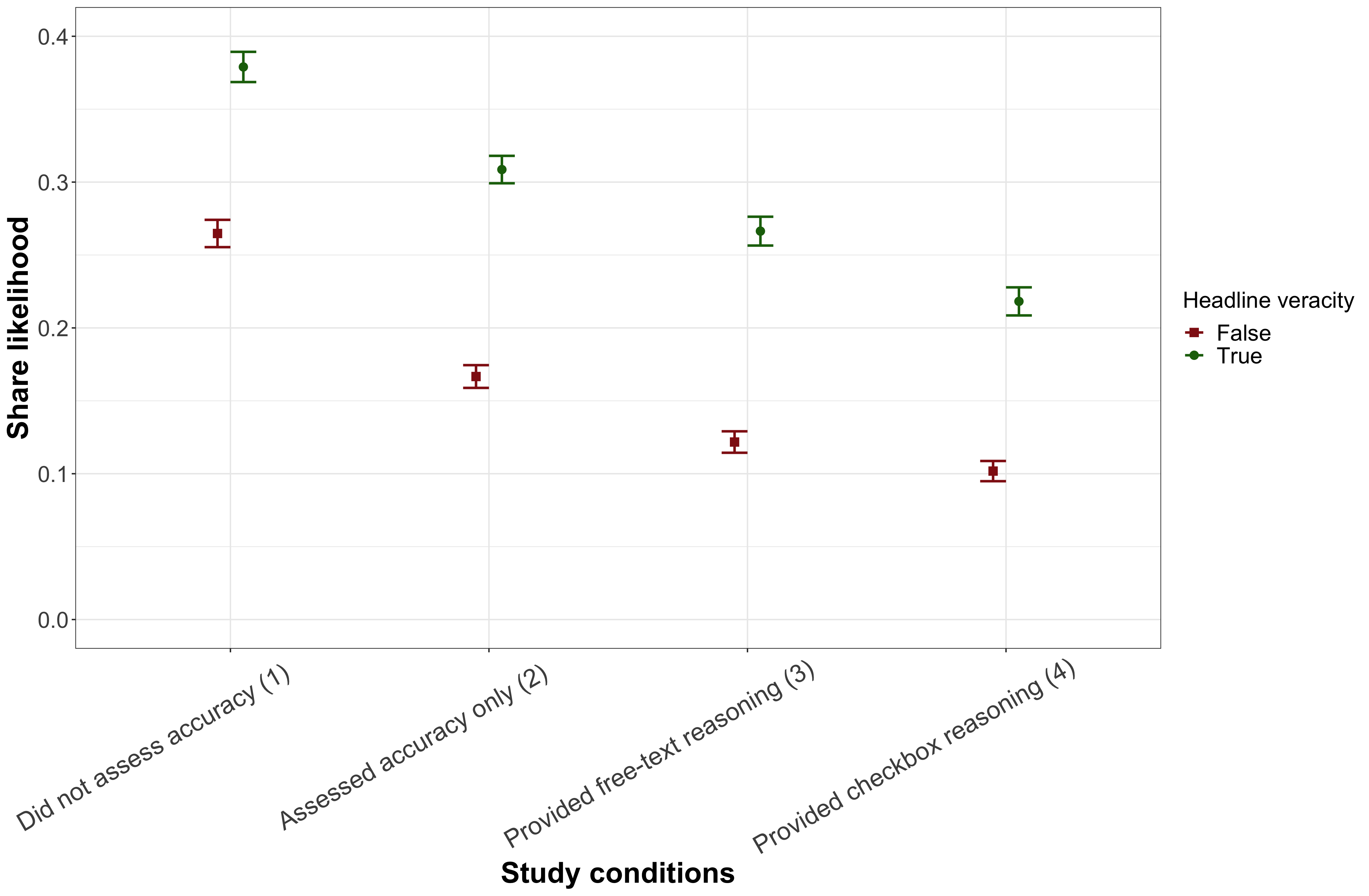}
 \caption{Share rate of true and false headlines across study conditions. The results suggest that people are less likely to share both accurate and inaccurate content if they are asked to assess the content's accuracy (condition 1 vs 2). We observe similar trends if they provide their reasoning in addition to assessing accuracy, compared with if they only assess accuracy (condition 2 vs 3). These interventions however, lower sharing of false content to a greater degree. The means of sharing true and false content also both decrease when people are asked to provide checkbox reasoning in addition to free-text (condition 3 vs 4). The ratio of shared false to true content however does not change.}
 \Description{}
 \label{fig:all_3}
\end{figure}

We observed that providing accuracy assessment had a significant effect on sharing intentions [$\chi^2(1)=38.05$, $p<0.001$]. Note that this variable was included in the veracity model only. Figure ~\ref{fig:all_3} shows sharing likelihood for condition 2, where we did request accuracy assessment, and condition 1, where we did not, across both true and false headlines. The results suggest that providing accuracy assessment about an article before deciding whether to share it lowers the probability that one shares the article for both true and false headlines. However, although this intervention results in a 18\% decrease in sharing of true headlines, the decrease in sharing false headlines is higher (37\%), therefore, reducing the ratio of false shared headlines to true ones.
The effect of the interaction between providing accuracy and veracity was not significant [$\chi^2(1)=3.22$, $p=0.07$]. Consistent with the results that follow, this finding can be because the headlines that participants perceived as accurate when they were prompted to deliberate on were a mix of objectively true and objectively false. Therefore, the sharing of both objectively true as well as false headlines was reduced. However, because sharing of objectively false headlines was less likely to begin with, the drop in sharing of false headlines was higher.

\paragraph{Effect of Providing Reasoning.}

We saw that whether participants provided reasoning had a significant effect on sharing intentions in both the veracity and perceived accuracy models [$\chi^2(1)=8.45$, $p=0.004$ for the veracity model, $\chi^2(1)=10.33$, $p=0.001$ for the perceived accuracy model]. 

Figure ~\ref{fig:all_3} shows sharing likelihood for condition 3, where we requested participants' rationale for why they believed a claim was or was not accurate, and condition 2, where we did not, across both true and false headlines.

Similar to the results we observed for requesting accuracy assessments, requesting reasoning reduces sharing of false headlines to a greater degree (27\%) compared to the decrease in sharing of true headlines (14\%). Therefore, the ratio of false shared headlines to shared true headlines is reduced.

\begin{figure}[!t]
 \centering
 \includegraphics[width=0.7\linewidth]{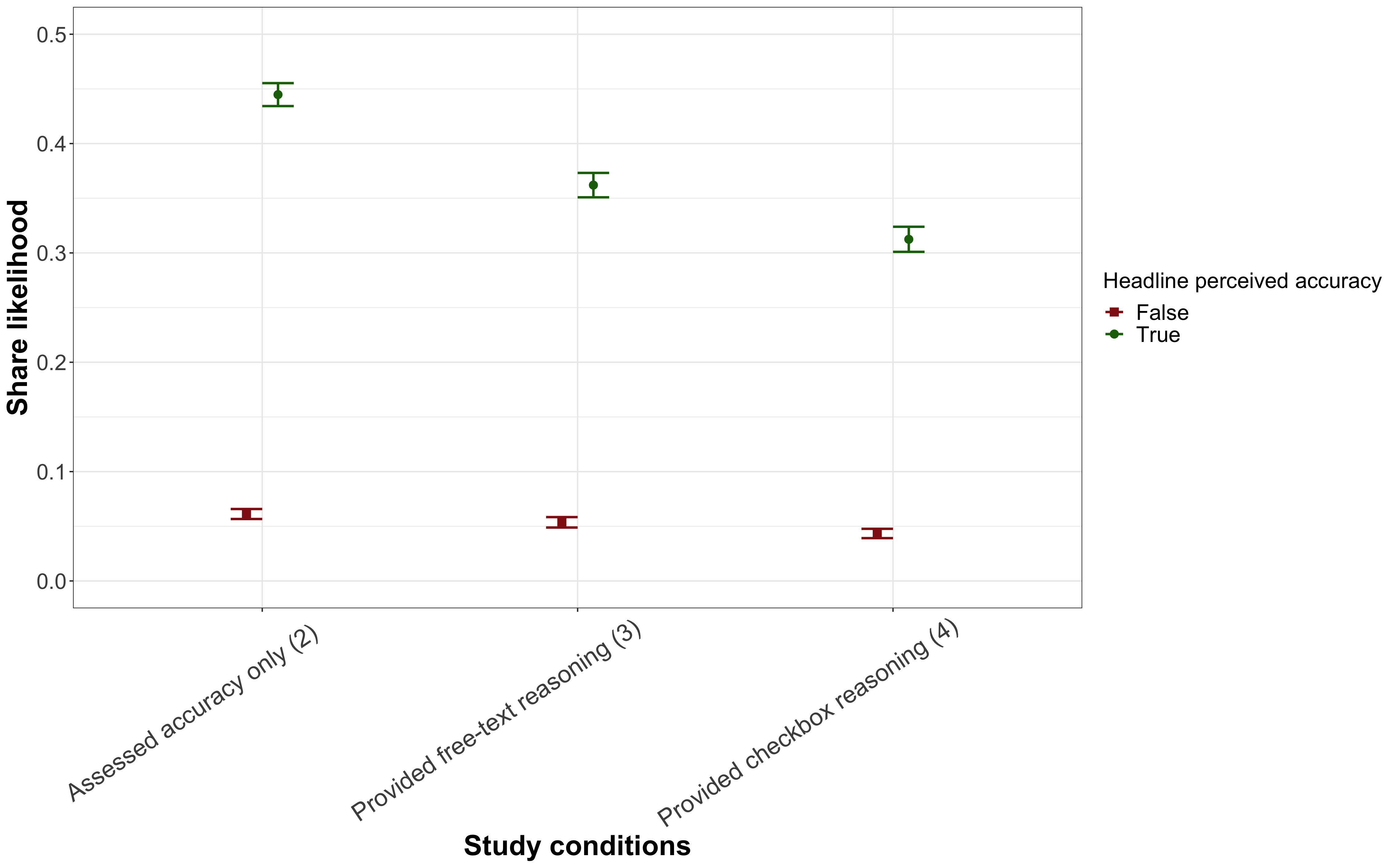}
 \caption{Share rate of headlines by their perceived accuracy, regardless of actual veracity. Participants are less likely to share content they initially perceived as true when they are asked about reasoning, or when they are requested to work through the checklist of reason categories. Sharing does not differ for headlines that were initially perceived as false, which could be because sharing of these headlines is rare to begin with.}
 \Description{}
 \label{fig:subjective_accuracy}
\end{figure}

Figure ~\ref{fig:subjective_accuracy} shows that requesting reasoning resulted in less sharing of headlines that participants initially believed as true but did not have an impact on sharing of perceived false content. The lack of reduction in sharing of subjectively false headlines however, could be because their sharing rate was very low to begin with and therefore there was not much room for improvement (6\% in condition 2, 5\% in condition 3). As expected, because the sharing of subjectively false headlines did not change to a great extent, but that sharing of headlines initially perceived as true decreased, the interaction between reasoning and perceived accuracy was significant in the perceived accuracy model [$\chi^2(1)=19.38$, $p<0.001$]. However, because headlines perceived as accurate were a mix of objectively true and objectively false, the sharing of both objectively true as well as false headlines was decreased (Figure~\ref{fig:all_3}). It is therefore reasonable that the interaction between reasoning and veracity was not significant in the veracity model [$\chi^2(1)=0.06$, $p=0.80$].

\paragraph{Effect of Reasoning Format.}

The effect of reasoning format on sharing was significant in both the veracity and perceived accuracy models [$\chi^2(1)=4.97$, $p=0.03$ for the veracity model, $\chi^2(1)=4.72$, $p=0.03$ for the perceived accuracy model]. 

Figure ~\ref{fig:all_3} shows that the sharing likelihood mean in the checkbox condition has a decrease of 17\% for false headlines, and 18\% for true headlines, compared to the free-text condition. Figure~\ref{fig:subjective_accuracy} shows that similar to the results we observed for requesting reasoning, presenting participants with reason checkboxes resulted in less sharing of content that participants initially perceived as true and did not lower sharing of headlines that were perceived as false. Sharing of content perceived as false however, was already rare (5\% in condition 3, 4\% in condition 4). 
This was the reason why we observed an interaction between reasoning format and perceived accuracy was significant in the perceived accuracy model [$\chi^2(1)=9.27$, $p=0.002$]. However, because headlines perceived as accurate were in fact a mix of objectively true and false, sharing of both objectively true and false headlines was reduced, so the interaction between reasoning format and veracity was not significant in the veracity model [$\chi^2(1)=3.48$, $p=0.06$].

\subsubsection{Reasons for Sharing Content Perceived as False.}

We examined participants' free-text responses to understand why they were willing to share a fraction, albeit small, of the headlines they perceived as false. One member of the research team used open coding to assign labels to participant responses (a total of 427). Of the responses that provided a reason, the most cited was that they believed the story was entertaining or that they thought it amusing to see which of their social media friends would believe the story: \textit{``If I felt in a playful mood I might post this just to see how many people no longer recognize satire''} (22\%). Others stated they would consider sharing the claim after fact-checking it: \textit{``If I could verify the contents this would be worth sharing.''} (20\%). Some considered the claim a debate starter: \textit{``I would share this because I think it would spark a good debate between pro and anti gun members. It would be interesting to see if people actually believe this information in the headline is true or not.''} (17\%). Other reasons included because they wanted to let their social circle know that the claim is false: \textit{``I would share this only to point out the misinformation available on most anything.''} (11\%) or that they wished to fact-check the claim: \textit{``to see if anyone can prove the authenticity of the image''} (10\%). Some participants believed it was important to inform their friends about the claim in case it turned out to be true: \textit{``Just in case it is accurate and, in either case, would make people look into the claim.''} (9\%). Some pointed out that they would share the article simply because it was interesting \textit{``It is interesting, even though I am not sure if it is true or not.''} (4\%). Others wished to share their emotions or frustration about the article with their social circle: \textit{``Just to point out how absurd the title of this article is.''} (4\%). 

Interestingly, in a few instances, we saw that although a participant had originally labelled a claim as inaccurate, they knowingly decided to share it to help advance their view: \textit{``Because I am against its [Marijuana's] legalization, maybe this would help instill fear in its users.''} (1\%) aligned with what was suggested in~\cite{marwick2018people}. In a few other occasions, we observed that participants had had a change of heart about the claim's accuracy: \textit{``I'd share because I know that it's not something that would just be out in the open like that and I know that he's stolen from the government by not paying his taxes, so I feel like it's accurate in a sense.''} (1\%), or that they wanted to provide an addendum to the claim to help rectify it: \textit{``So I could type. "...Inadvertently?" (cough) Because Facebook is evil, not Cenobite evil, but corporation-evil. They did that on purpose, they know it, and I know it. The trouble is how few OTHER people know it.''}---\textsf{Claim: Some Phone Cameras Inadvertently Opened While Users Scrolled Facebook App.}

\subsubsection{Factors Interacting with Findings}

\paragraph{Confidence.}

\begin{figure}[htbp]
 \centering
 \includegraphics[width=\linewidth]{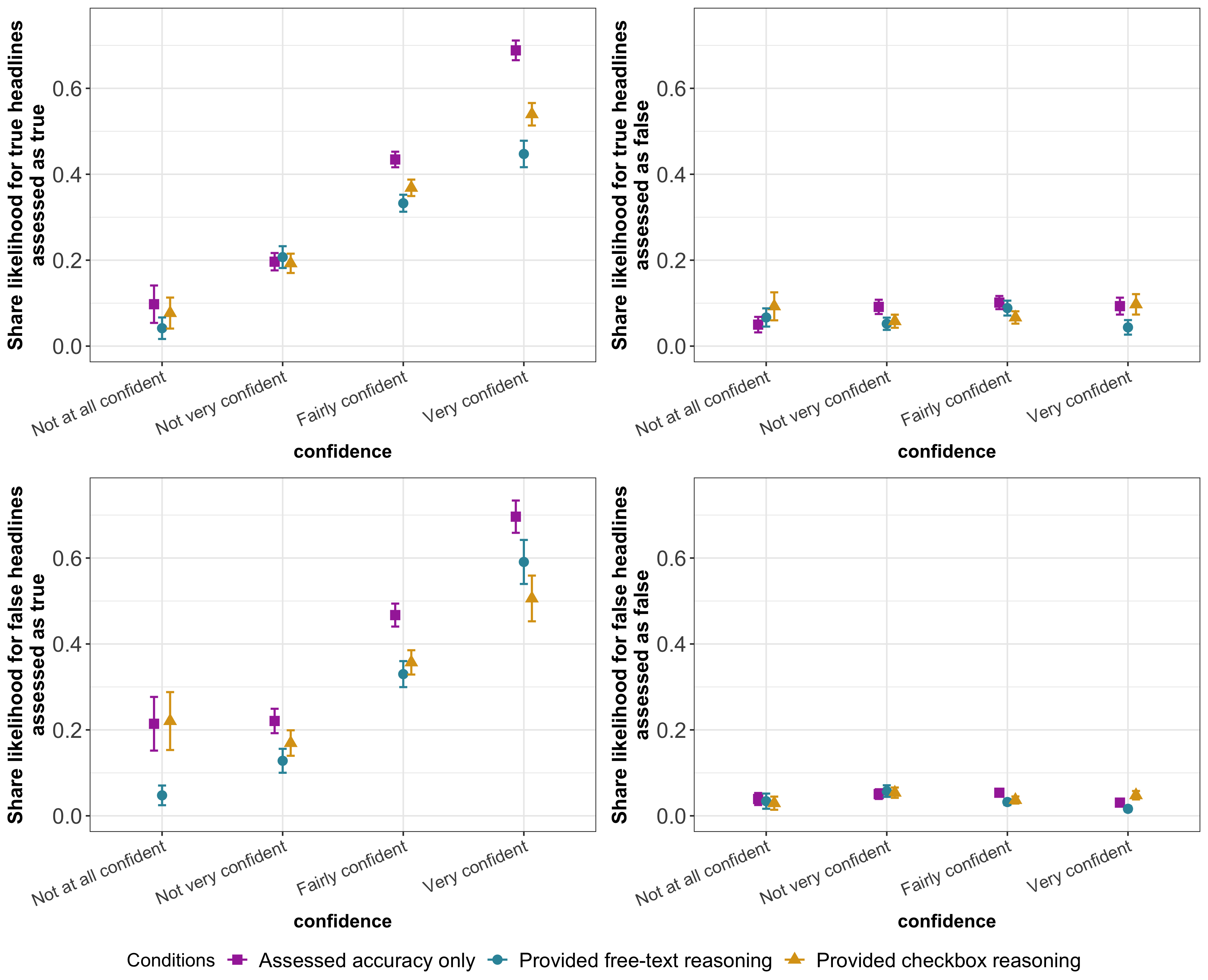}
 \caption{Share rate of true and false headlines across different confidence levels. For true headlines that they correctly assess as true, participants are less likely to share content they are less confident about regardless of whether they are asked for their reasoning. It is on those headlines about which they are more confident that requesting reasoning plays a role. For false headlines that they mistakenly assess as true, asking about reasoning plays a role in sharing across all confidence levels.}
 \Description{}
 \label{fig:confidence}
\end{figure}

We observed that asking people to provide reasoning inhibits sharing of true as well as false content. We wished to see how sharing behavior across our treatments differs with regards to how confident participants in each condition are in their accuracy assessments. For instance, it is possible that asking people for their rationales makes them reluctant to share headlines on whose accuracy they report lower levels of confidence. Figure ~\ref{fig:confidence} shows that participants are not likely to share headlines that they assess as false regardless of whether they are confident about their assessment. For headlines that they assess as true however, in all the treatment conditions, they are less likely to share headlines they are less confident about.

We expect that the participants who are asked to provide accuracy assessments or reasons will be less willing to share headlines about which they initially are more confident, compared to those participants who are not. This is what we observe for false headlines that participants initially misjudged as true. However, surprisingly, for true headlines that participants had correctly judged as true, the intervention does not reduce sharing at lower confidence levels. It is on those headlines about which participants report higher confidence that the reasoning treatment and the reason checkboxes play a role.

Note that we asked participants to indicate their level of confidence before they provided their reasons and they could not return to the confidence question once they advanced to the question requesting reasoning. It is possible that reflecting on their rationale has lowered their confidence.

\paragraph{Average Perceived Accuracy.}

It is conceivable that there are headlines that are in fact accurate but sound too outlandish to be true. And conversely, actual false headlines can seem reasonable. While we found that headline veracity does indeed have a strong effect on whether people are willing to share the headline, we wanted to tease apart the ground truth from how accurate a claim was perceived as according to the wisdom of the crowd. Therefore, we assigned to each headline an \textit{average perceived accuracy} metric, which was the average of accuracy assessments from all the participants who had provided accuracy assessments for the headline, mapping accurate to 1, and inaccurate to 0. With this metric, a headline would no longer have a dichotomous ground truth, and instead, would have a degree of truth.

Overall, actual true headlines had a higher average perceived accuracy compared to the false ones. Table~\ref{tab:perceived_accuracy_examples} presents examples of headlines that were judged by most participants correctly or incorrectly, and some in between.

\begin{footnotesize}
\begin{table*}[!t]
\setlength{\aboverulesep}{0pt}
\setlength{\belowrulesep}{0pt}
\caption{Examples of easy, medium, and hard calls around headline accuracy assessed by the perceived accuracy of the headline averaged over all participants who assessed the headline. The average perceived accuracy is on a scale of 0-1, with 0 indicating that the headline was perceived as inaccurate and 1, as accurate.} 
\label{tab:perceived_accuracy_examples}
\centering
\begin{tabular}{cp{7.5cm}cc}
 \toprule
 \rowcolor{gray!16} Difficulty & Headline & Veracity & Avg. perceived accuracy  \\
 \midrule
    \cellcolor{gray!16} & A Study Showed That Dogs Exhibit Jealousy & True & 0.90 \\
    \cline{2-4}
    \multirow{-2}{*}{\cellcolor{gray!16} Easy} & Sipping Water Every 15 Minutes Will Prevent a Coronavirus Infection & False & 0.05\\
    \midrule
    
    \cellcolor{gray!16} & President Trump's Awarding of a Purple Heart to a Wounded Vet Went Unreported by News Media & True & 0.49 \\
    \cline{2-4}
    \multirow{-2}{*}{\cellcolor{gray!16} Medium} & Eric Trump Tweeted About Iran Strike Before It Was Made Public & False & 0.45\\
    \midrule

    \cellcolor{gray!16} & There Are More Trees on Earth Than Stars in the Milky Way & True & 0.25 \\
    \cline{2-4}
    \multirow{-2}{*}{\cellcolor{gray!16} Hard} & Rain That Falls in Smoky Areas After a Wildfire Is Likely to Be Extremely Toxic & False & 0.62\\
 \bottomrule
\end{tabular}
\end{table*}%
\end{footnotesize}

We built a linear model to explain share likelihood of a headline predicted by its average perceived accuracy. The dependent variable of the model was the average of share intentions for a headline from all the participants that had been presented the headline, mapping the 3-item Likert outcomes ``Yes'', ``Maybe'', and ``No'' to numeric values as explained in~\ref{section:nudge_main_results}. The independent variable was the headline's average perceived accuracy (continuous). We fit this model to the data from each of the control and treatment conditions separately. Because the average perceived accuracy of each headline was calculated over the whole dataset, it remained constant across conditions. Share average however, varied in each condition.

\begin{figure}[htbp]
 \centering
 \includegraphics[width=\linewidth]{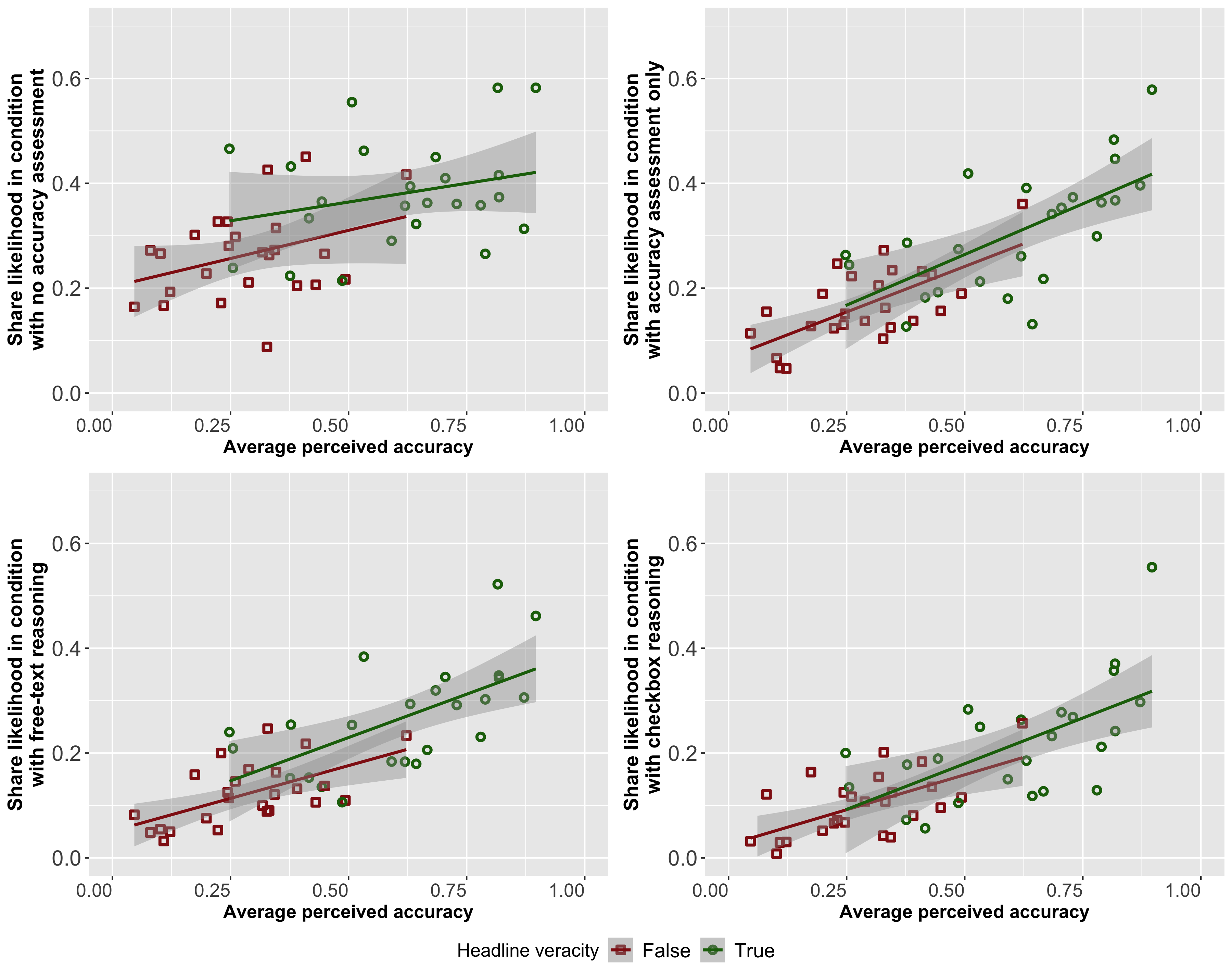}
 \caption{Share likelihood of each headline by its perceived accuracy averaged over all treatment conditions. The slopes in the treatment groups are higher than the control, suggesting more sharing differentiation between headlines that were perceived as true vs false. The fitted lines in the treatment groups also have narrower confidence intervals, suggesting less dispersion and uncertainty in sharing intentions. }
 \Description{}
 \label{fig:truthiness}
\end{figure}

Figure ~\ref{fig:truthiness} shows how participants' sharing intentions differ across conditions as the average perceived accuracy of a headline increases. In the treatment groups where we asked for accuracy assessment or reasoning in addition to accuracy, the slopes are higher compared to the control condition. This observation suggests that the interventions helped people be more differentiating in sharing of headlines that they perceived as true vs false. The confidence intervals around the mean also seem to be narrower for the treatment conditions compared to the control. With the numbers of participants across treatment conditions being almost similar to or less than that of the control, a narrower confidence interval around the treatment slopes suggests less dispersion and uncertainty in sharing intentions.

\paragraph{Demographics.}
We conducted an exploratory analysis of the demographics of participants and their effect on share likelihood. We have reported these analyses in Appendix~\ref{section:nudge_demographics}.

\subsection{Discussion of Nudge Study Findings}

This study investigated the effects of two types of interventions on people's intention of sharing headlines. The interventions included asking people to assess the accuracy of the headline and to provide their rationale for why they believe the headline is or is not accurate. We observed that the participants who assessed the accuracy of a headline were less likely to indicate they are willing to share it, compared to those participants whom we did not prompt about accuracy. Although the intervention resulted in curbing of sharing both true and false content, the reduction in sharing of false headlines was higher. Our results corroborate prior findings that nudging people to be mindful of news accuracy increases sharing discernment \cite{pennycook2020fighting}, and enrich our understanding of how impactful different nudges will be.

We found that asking people to reflect and elaborate on the reason for a headline's (in)accuracy further lowers the probability that they share either true or false content, compared to if they only assess the headline's accuracy. The intervention however, reduces sharing of false content to a greater degree.

We observed that the decrease in sharing as a result of requesting people to provide structured reasoning was almost similar across true and false content (17\% for false, 18\% for true in our sample). Although this observation suggests that selecting from a checklist of rationales in addition to providing accuracy assessments and free-text reasoning does not seem to help in further differentiating between false and true content, such a checklist can still be used on platforms for the added benefit of capturing and surfacing reasons in structured form.

In addition, we examined how sharing intentions differ across conditions as the average perceived accuracy of headlines increases. We observed that the interventions caused people to be more differentiating in sharing of content that they perceived as true vs false compared to the control condition and that the dispersion of sharing decisions in the treatment groups was also lower. Interestingly, Figure ~\ref{fig:truthiness} shows that the slopes of sharing by perceived accuracy across all conditions are less than 1, indicating that as the perceived accuracy of a headline increases, its share likelihood also increases but at a lower rate. One possible explanation for this phenomenon is that headlines that most people agree are true could appear as less interesting and already believed to be known by the others in one's social circle.

One concern around the generalizability of our results to online platforms is the possible existence of Hawthorne effect, under which participants change their behavior due to an awareness of being studied \cite{mccambridge2014systematic, kreuter2008social, preist2014competing}. We therefore need to understand if participants would exhibit the same behavior as observed in our study if they were placed in a different intervention condition. In a user study with users recruited from worker platforms, Pennycook et al. investigated how self-reported share likelihood of headlines were influenced by an accuracy nudge at the onset of the study where as a pre-task they asked users to assess the accuracy of a single news item. They found that the participants' likelihood of sharing false headlines compared to true decreased with the intervention, similar to what we observed in our study. However, they reported that asking users to instead assess the humorousness of a headline did not yield similar results. In another study Pennycook et al. sent an unsolicited message to Twitter users who had recently shared links to websites that produced false or hyperpartisan content and asked them to assess the accuracy of a non-political headline. They report that the quality of the news sites that these users shared in the 24 hours after receiving the message was increased compared to the sites shared by others who had not received the message \cite{pennycook2021shifting}, suggesting that the effects of these behavioral nudges will indeed generalize to social media platforms.

In addition, other prior work has reported that there is a correlation between hypothetical sharing of news stories reported by survey respondents and actual sharing on social media ~\cite{mosleh2020self}. With our nudges proving to be effective and the results generalizing to actual social media platforms, platform designers can implement these nudges to encourage more informed propagation and consumption of news.

As we discuss in Appendix~\ref{section_potential_confounds}, the difference in spam rates across conditions may have influenced the makeup of the data. However, if these interventions are implemented on social media, it is unlikely that we will observe similar spamming behaviors as we do on a paid worker platform. Spamming in lieu of providing legitimate rationales by itself could be a clear signal of the sharer's credibility to those who will encounter the shared news.

It is clear that other factors beyond the perceived accuracy of the headline can affect sharing intentions.
Some participant responses indicated that they were reluctant to share a headline because they were not interested in the topic---\textit{``It's not a story that I would share with others, I'm not interested and my followers wouldn't be either.''}, the headline referred to a sensitive topic and might create controversies---\textit{``It is a sensitive topic that I feel strongly one way about and I don't want to start disagreements on my page.''}, they simply do not share on social media---\textit{``I don't share anything on social media.''}, they did not want to overburden their social media friends with information---\textit{``I would feel bad clogging up peoples timelines with useless information that is already well known.''}, even though the claim was true, it shed a good light on someone of the opposite party---\textit{``It is positive about Trump. I would never post anything positive about him.''}, or even though the claim was true, it put someone affiliated with their party in a bad light---\textit{``I wouldn't bad mouth Joe Biden''}, aligned with what is reported in prior work~\cite{shin2017partisan}. Nevertheless, the primary cited reason for not sharing a headline was that it was of dubious credibility---\textit{``That is just an outright lie. I refuse to contribute to the misinformation being shared on Facebook.''}.

Therefore, sharing intentions, or lack thereof, can be used as a proxy to gain insight into how accurate participants believed each headline to be post interventions. Although we did request accuracy assessments across the two treatment groups where participants provided reasoning, these accuracy assessments were done at the onset of each item and before participants were subjected to the reasoning interventions. In addition, the control group did not provide accuracy assessments. It is by contrasting the sharing intentions and not accuracy assessments that we understand the effects of our interventions. Figure ~\ref{fig:truthiness} confirms this paradigm that share probability increases with the perceived accuracy of a claim.

\section{Discussion}

We have discussed the two studies in their respective sections; here we offer more general observations.

The behavioral nudges that we tested in this study, providing accuracy assessment and rationale for whether a news story is or is not accurate, proved to be effective at reducing the ratio of false content to true that participants indicated they were willing to share. The reduction of shared false headlines came at the cost of also curbing the sharing of true ones. However, we still believe that the outcome is preferable to leaving misinformation unchecked and unchallenged.
Indeed, it is conceivable that platforms can benefit from an overall improved engagement if the feeds that they present to their users become more reliable. In addition, even if approaches similar to the nudges in our study result in loss of some profit, the implications that unmuddied online information spaces have for the society may warrant persuading platforms, through activism or legislation, to adopt them.

\subsection{Design Implication}

While platform moderators and fact-checkers play a valuable role in flagging and removing content that has already spread and become visible, other measures are needed to restrain sharing of misinformation as it is being handed from user to user. 
In addition, although policy-driven platform moderation is necessary in some contexts, communities should be wary of relinquishing all the power of content filtering and highlighting to the platforms whose incentives as for-profit entities running on ads do not necessarily align with the users’ \cite{grygiel2019social, Facebookwarren}. The challenge of moderation is exacerbated as not all accounts of problematic behavior or posts can be provisioned a priori in platform policies, leading moderators to make ad-hoc decisions in grey areas that sometimes draw criticism~\cite{Facebooknapalm, ibrahim2017facebook, Facebookkurdish, FacebookPragerU}.

These challenges suggest that the problem of misinformation could additionally be tackled at the user level. The interventions that we studied in this work aim at this problem by first, introducing a barrier, albeit low, to posting and sharing, and second, shifting users' attention to accuracy and away from the social feedback and engagement that they would receive at posting time.
Our interventions can be used in the existing social media platforms such as Twitter and Facebook and are aligned with the initiatives they have already undertaken to combat the proliferation of misinformation. However, the usefulness of these nudges is not limited to these platforms as they can be leveraged in alternative platforms with different publishing models, such as WikiTribune where users curate content collectively~\cite{o2019you}.

We envision that requesting reasoning on social media or content sharing platforms can be done in a similar fashion to how emotions and reactions are currently captured on the existing social media or how users cite references when developing content on wiki-based platforms.
Structured reasons provided by users for or against a post's accuracy can serve as rich metadata based on which other users can filter the posts they would want to view. In such a scenario, a user might choose to view only those articles that have been evaluated as true by a friend because the evaluator has asserted they have domain knowledge on the subject. The taxonomy that we developed originates from people untrained in credibility signifiers often developed by experts, and we determined in our studies that other untrained people can reliably use it to provide their rationales. Therefore, the adoption of requesting rationales via a checklist similar to that of our study on social media does not appear to incur a barrier to entry for users, except maybe in forcing some degree of deliberation which is desirable. Prior work reports that the effectiveness of fact-checking depends on the relationship of the user offering the fact-checking information with the user requesting it or one who has produced an inaccurate post \cite{margolin2018political, hannak2014get}. Therefore, by incorporating accuracy assessments of our study in platforms and making them visible and accessible to users, we hope social media friends can benefit from them.

\section{Future Work}

One direction for future work is to examine how users react to posts accompanied by such rationale tags as the ones in our study and what factors they consider when deciding the credibility of a post or the persuasiveness of its tagged rationales. However, because who the sharer of a post is can also impact perceived content credibility \cite{flintham2018falling}, such a study would be more informative if conducted as a field study on a social network, rather than a controlled experiment.

Our work examined the effects of accuracy assessment and reasoning nudges on content sharing when users are required to provide them. Future work can investigate the effects of allowing users to optionally provide these signals, similar to how ``likes'' and ``upvotes'' are captured on social media.

Although one reason why people would share a story was to inform the others, we found that there are various reasons they might share even a headline they do not necessarily believe to be accurate. Such reasons include because the article is entertaining or to ask their social circle to help them with fact-checking the story. Future work could investigate how providing a checklist of these sharing intentions in a fashion similar to our study would affect sharing behaviors on social media and how it would impact the consumption of posts on which these signals are provided by the users who encounter them.

One interesting observation from our exploratory analyses in Appendix ~\ref{section:nudge_demographics} was that our Republican participants were more likely to share false claims compared to our Democratic participants. While our analyses were not planned a priori, they are bolstered by prior studies that have reported similar findings~\cite{guess2019less}. These observations give rise to interesting questions that could be examined in future work, such as whether behavioral nudges should be applied indiscriminately to all platform users or only those who have been found to habitually share misinformation, or whether users should be primed every time they intend to share posts or if it is sufficient to apply the nudges only occasionally.

\section{Limitations}
In condition 4 of the Taxonomy study where we presented the checklist of reason categories, we also asked participants to explain their choices in free-text to examine how they had used the categories and if they had truly understood their intended use. With this model, comparing this condition with condition 3 where participants provided their reasoning via free-text gives us insight into whether restricting people's rationales to the taxonomy framework works as well as otherwise allowing them to provide unstructured reasons. However, our study did not include another condition where participants needed to only work through the checklist without elaborating on their choices. The inclusion of such a condition would have allowed us to determine whether the checklist of reasons can replace free-text entirely and induce the same level of discernment. Not requiring free-text but rather providing it as optional in addition to the checklist could potentially be more desirable for the adoption of this strategy on social media.
Future work can include a condition of this nature.

\section{Conclusion}

In this work, we explored how to alter social media platforms such that users would have content accuracy in mind upon sharing posts. We explored nudges that could be used at scale intended to encourage users to think whether a post is accurate and their rationales for believing so. To facilitate capturing people's rationales in a structured format and help the adoption of the nudges on social media, we conducted a study where we developed a taxonomy of reasons why people believe or disbelieve news claims.
That study involved presenting news claims to people as well as the taxonomy and asking them to use the reason categories to provide their rationales for the (in)accuracy of the claims. We conducted multiple iterations of the study while revising the taxonomy until participants could reliably use it to label their responses.

We then examined the effects of two different nudges, accuracy assessment and providing reasoning for why a news story is or is not accurate, on people's sharing intentions on social media. We found that both nudges reduce sharing of true and false content, but the decrease in sharing of false content was higher. Our findings on the effects of the accuracy and reasoning nudges offer implications for social media platform designers on how to mitigate sharing of false information. Furthermore, these platforms can ask their users to provide accuracy assessments for the posts the users share by guiding them through the taxonomy categories that we developed in the study. These structured reasons could potentially help those who encounter the post e.g., by enabling them to filter their newsfeed based on different reasons that post sharers have specified, arguing for or against a post's accuracy.

\section{Acknowledgments}
We would like to thank Ezra Karger, Ali Kheradmand, and Mohammad Amin Nabian for their valuable feedback regarding the statistical analyses.

\bibliographystyle{ACM-Reference-Format}
\bibliography{bibliography}

\appendix

\section{Association Between Participants' Demographics and the Taxonomy Categories.}
\label{section:demographics_taxonomy_study}

We performed an exploratory analysis on the data from our Taxonomy study to understand whether the demographics of participants influence the types of rationales they give for why they believe a claim is or is not accurate. This analysis was not part of our study design and was added as a stepping stone for future work. We limited our analyses to the data obtained from the last iteration of the study because the categories that the participants and the research team coder had used in the prior iterations had changed. We further excluded those datapoints for which we did not have the gender, party, and ethnicity of the participant, resulting in 953 datapoints of which 645 had free-text elaboration (did not belong to the \textit{Don't know} categories). 

For party, each participant was labeled either a Democratic or a Republican. We were able to place all participants including those who identified as Independent or other (e.g., Green) in one of the two Democratic or Republican categories because in addition to party, we had asked participants about their political preference (strongly Republican, lean Republican, Republican, Democrat, lean Democrat, strongly Democrat). Because the majority of our participants were White, ethnicity was given the values of White, and Not White. With respect to education, we categorized the participants as having a college degree including Associate's degree vs not. 

We then performed Chi-square tests of independence on the contingency tables of rationales and each of the demographic factors of party, gender, and ethnicity. We caution that these tests were underpowered considering the number of categories and our sample size and future studies are needed to ascertain whether our results hold true with a larger sample. 
The tests did not find a statistically significant association between rationales that participants gave and their party or ethnicity [$\chi^2(22)=30.03$, $p=0.12$ for party; $\chi^2(22)=21.80$, $p=0.47$ for ethnicity]. However, the rationales were not independent of the participants' gender [$\chi^2(22)=40.87$, $p=0.009$].

\begin{figure}[!t]
\centering
\begin{minipage}{\textwidth}
 \centering
 \includegraphics[width=\linewidth]{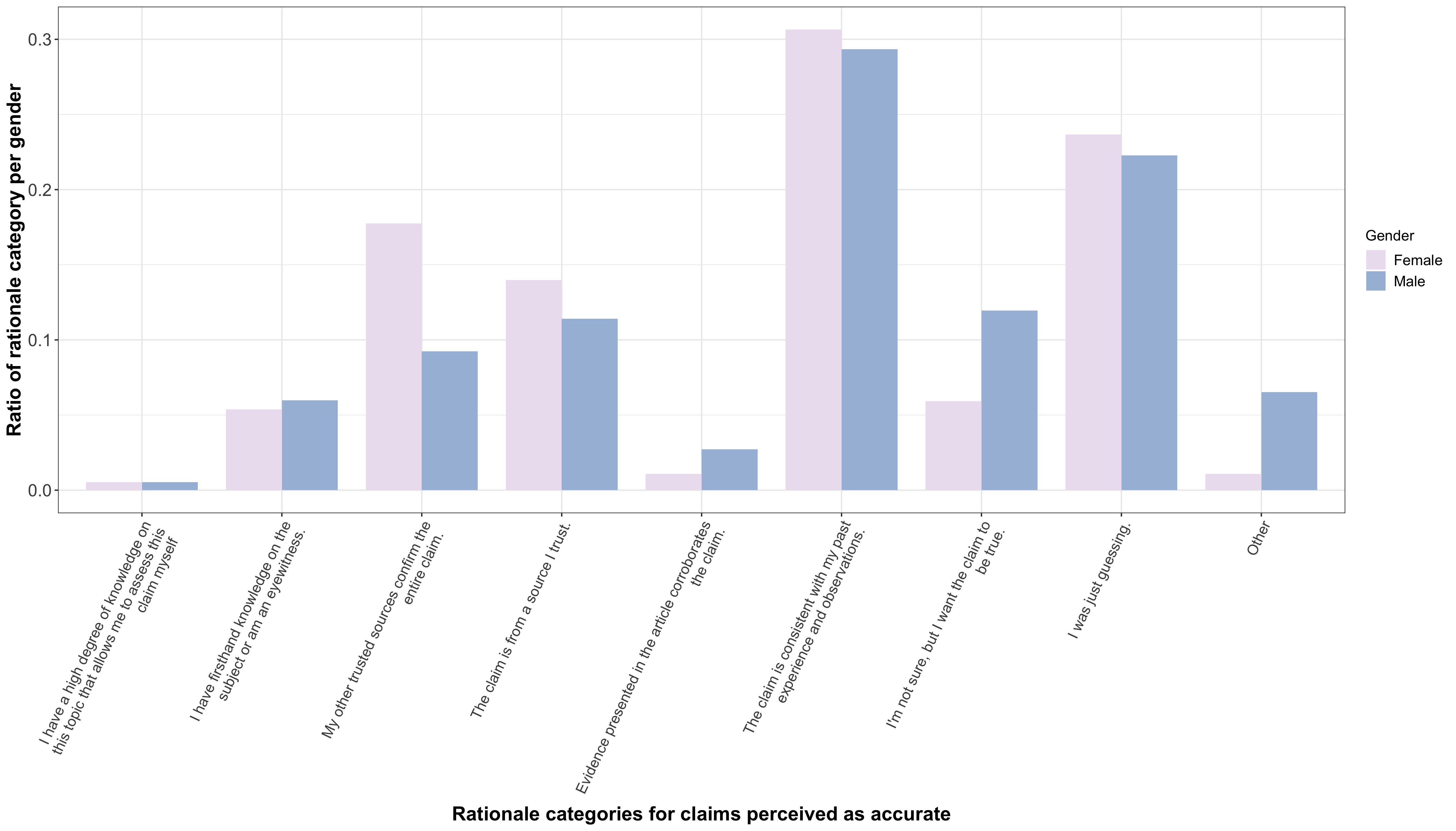}
\end{minipage}
\begin{minipage}{\textwidth}
 \centering
 \includegraphics[width=\linewidth]{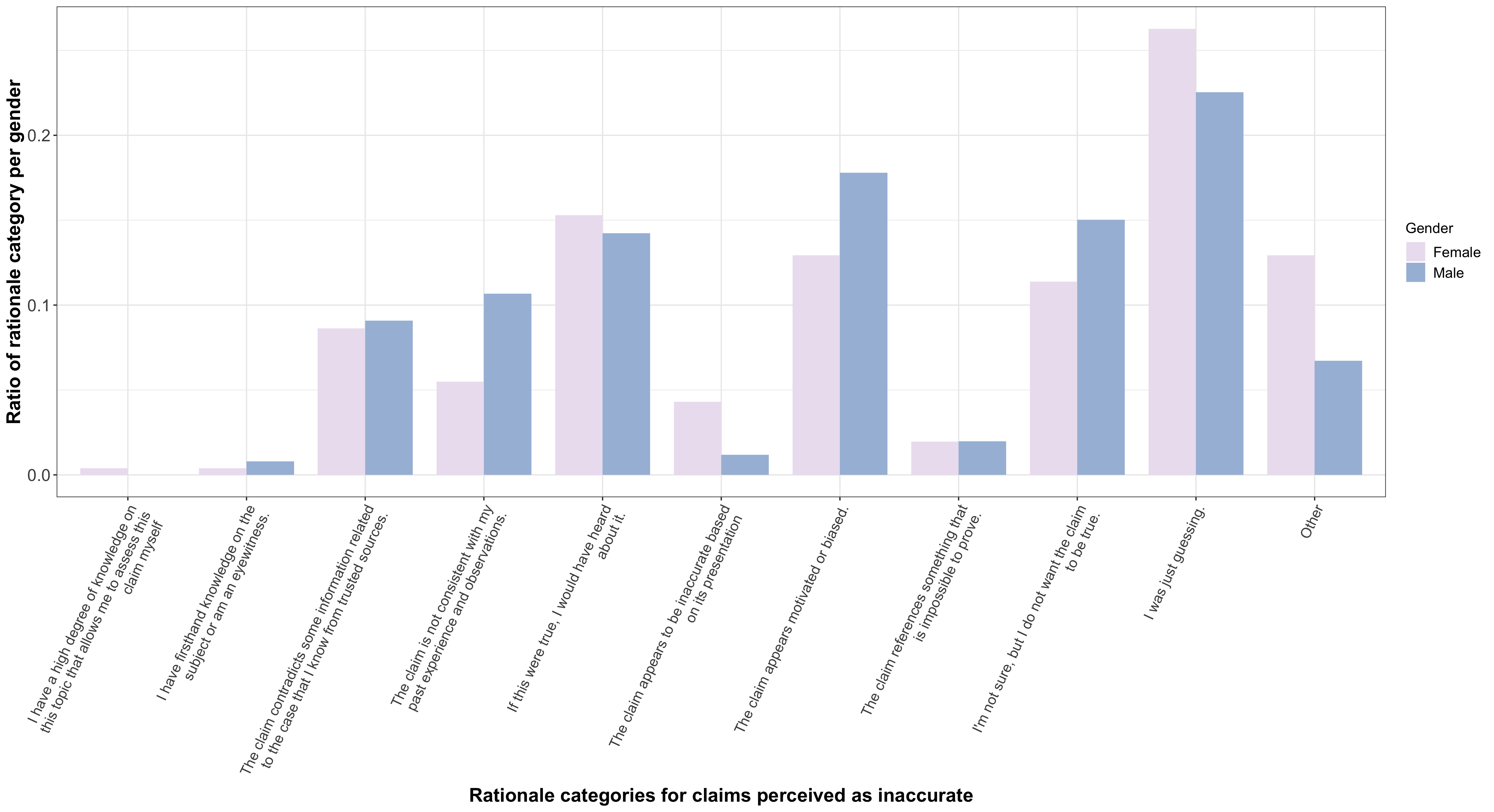}
\end{minipage}
\caption{Distributions of rationales by the gender of the user for claims perceived as accurate and inaccurate. Each bar shows the ratio of the rationale category relative to all rationales asked by users of the same gender.}
\label{fig:accurate_inaccurate_by_gender}
\end{figure}

\begin{figure}[t]
 \centering
 \includegraphics[width=0.5\linewidth]{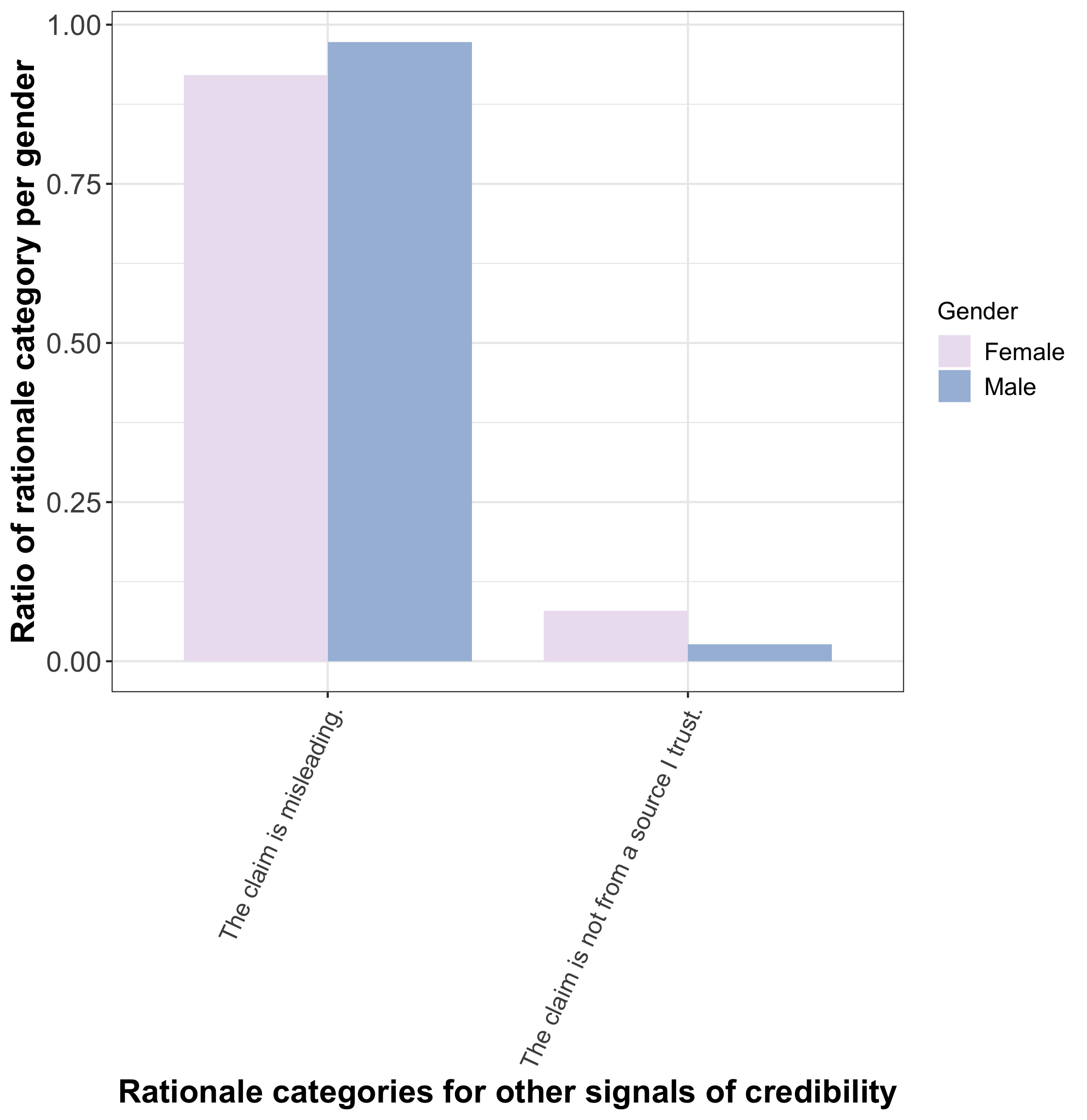}
 \caption{Distributions of rationales by the gender of the user for other signals of credibility besides accuracy. Each bar shows the ratio of the rationale category relative to all rationales asked by users of the same gender.}
 \Description{}
 \label{fig:other_signals_by_gender}
\end{figure}

Figures~\ref{fig:accurate_inaccurate_by_gender} and ~\ref{fig:other_signals_by_gender} show how the distributions of rationales across categories vary by the user's gender. Because the numbers of rationales provided by each gender is different, the bar for each question category and each gender is normalized by the number of all questions asked by users of the same gender.

\section{Effects of the Demographics of the Nudge Study Participants} 
\label{section:nudge_demographics}

We performed an exploratory analysis of the demographics of the Nudge study participants to understand what role these factors play in participants' decision of whether to share headlines. We had not planned for these analyses a priori in our experiment design, but later included them for completeness. Thus, this analysis should be considered exploratory, and $p$-values not indicative of true significance. We developed a linear model with share intention as the dependent variable, with outcomes ``Yes'', ``Maybe'', and ``No'' mapped to numeric values as explained in 6.2, in addition to several demographics factors:
\begin{multline}
  \text{share} \sim \text{veracity} \times (\text{partisanship concordance} \times (\text{accuracy condition} + \text{reasoning condition} + \\\text{reasoning format)} + \text{party} + \text{gender} + \text{age} + \text{ethnicity} + \text{education}) + (1|\text{participant}) + (1|\text{claim})
\end{multline}
Similar to the model in~\ref{section:nudge_main_results}, we included the veracity of the headline and treatment conditions as independent variables, and claim and participant as random effects. We limited this analysis to that portion of the data for which we had the complete demographics information required for our model, excluding 638 datapoints. In addition, we excluded 46 datapoints from the participants who had identified as neither male nor female because these datapoints were too few for fitting the model.

We treated party, ethnicity, and education similar to Appendix ~\ref{section:demographics_taxonomy_study}. We binned age into 7 buckets. We did not include the interaction between the demographic factors because given our sample size, we did not have enough power to do so.

In the model, we also included \textit{partisanship concordance} which was a measure of the alignment between the participants' self-declared party and the partisanship rating that they had given to the headline (measured on a 5-item Likert scale). This value ranged from 1-5 with 1 indicating no alignment, and 5 complete alignment.
\begin{multline}
\text{concordance} = (\text{partisanship rating of the headline}) \times (\text{rater party} == \text{Republican}) + \\(6 - \text{partisanship rating of the headline}) \times (\text{rater party} == \text{Democratic}) 
\end{multline}

Because this model includes several demographic factors that may capture some degree of variance in share likelihood, the effects of the treatments observed in this more refined model can serve as a confirmation of the results in~\ref{section:nudge_main_results}. The results obtained for the demographic factors however, should be taken with caution and further examined in future work, as these were not planned analyses.

We performed a Wald Chi-Square test on the fitted model to determine which of the factors had a significant effect. Consistent with the results in~~\ref{section:nudge_main_results}, the effects of veracity, providing accuracy, providing reasoning, and reasoning format were significant [$\chi^2(1)=33.04$, $p<0.001$ for veracity, $\chi^2(1)=33.62$, $p<0.001$ for providing accuracy, $\chi^2(1)=7.59$, $p<0.01$ for providing reasoning, $\chi^2(1)=4.88$, $p=0.03$ for reasoning format]. The interaction between veracity and whether the participant was asked about accuracy was also significant at the $\alpha=0.05$ level [$\chi^2(1)=4.15$, $p=0.04$].
The sample means shown in Figure~\ref{fig:all_3} for conditions 1 and 2, suggest that although accuracy assessment reduces sharing of both false and true content, when users are asked to assess accuracy, the reduction in sharing affects false headlines more compared to true headlines.  

In addition, we observed that the effects of a number of demographic factors were significant as well.

\begin{figure}[!t]
\centering
 \begin{minipage}{\textwidth}
 \centering
 \includegraphics[width=0.45\linewidth]{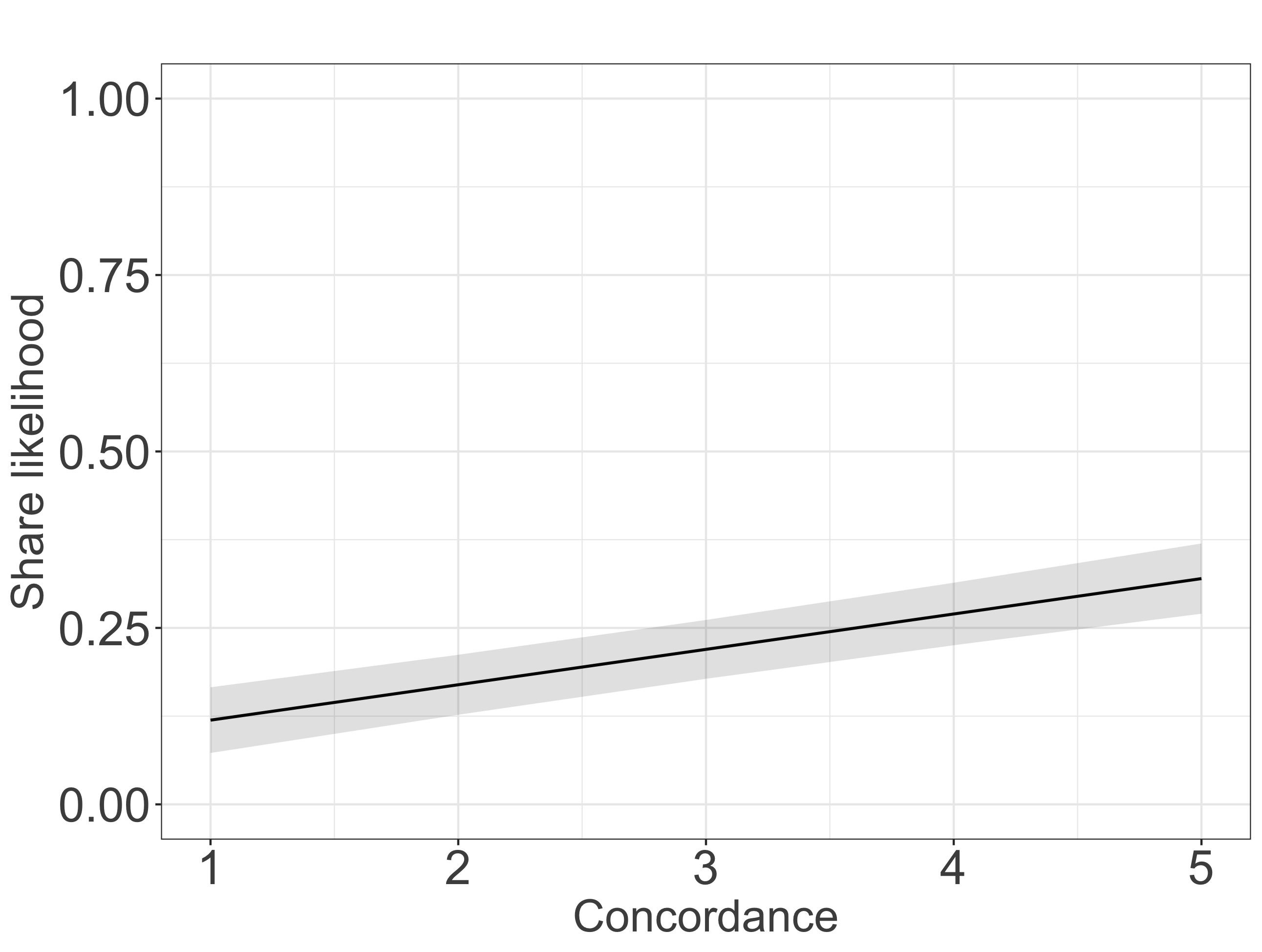}
 \captionof{figure}{Predicted values (marginal effects) for share likelihood as concordance (alignment between headline and participant partisanship) increases obtained from the model with demographics included as independent variables.}
 \label{fig:concordance}
\end{minipage}
\end{figure}

\paragraph{Concordance} had a statistically significant effect on sharing intentions [$\chi^2(1)=256.52$, $p<0.001$]. Figure~\ref{fig:concordance} displays the predicted values (marginal effects) for share likelihood as concordance increases. As the alignment between a participant's party and their perceived partisanship of a headline increases, the probability that they share the headline increase as well. This observation aligns with prior studies that have found people are more likely to consider sharing politically concordant headlines than discordant headlines \cite{pennycook2021shifting, shin2017partisan}.

\begin{figure}[!t]
\centering
\begin{minipage}{.45\textwidth}
 \centering
 \includegraphics[width=\linewidth]{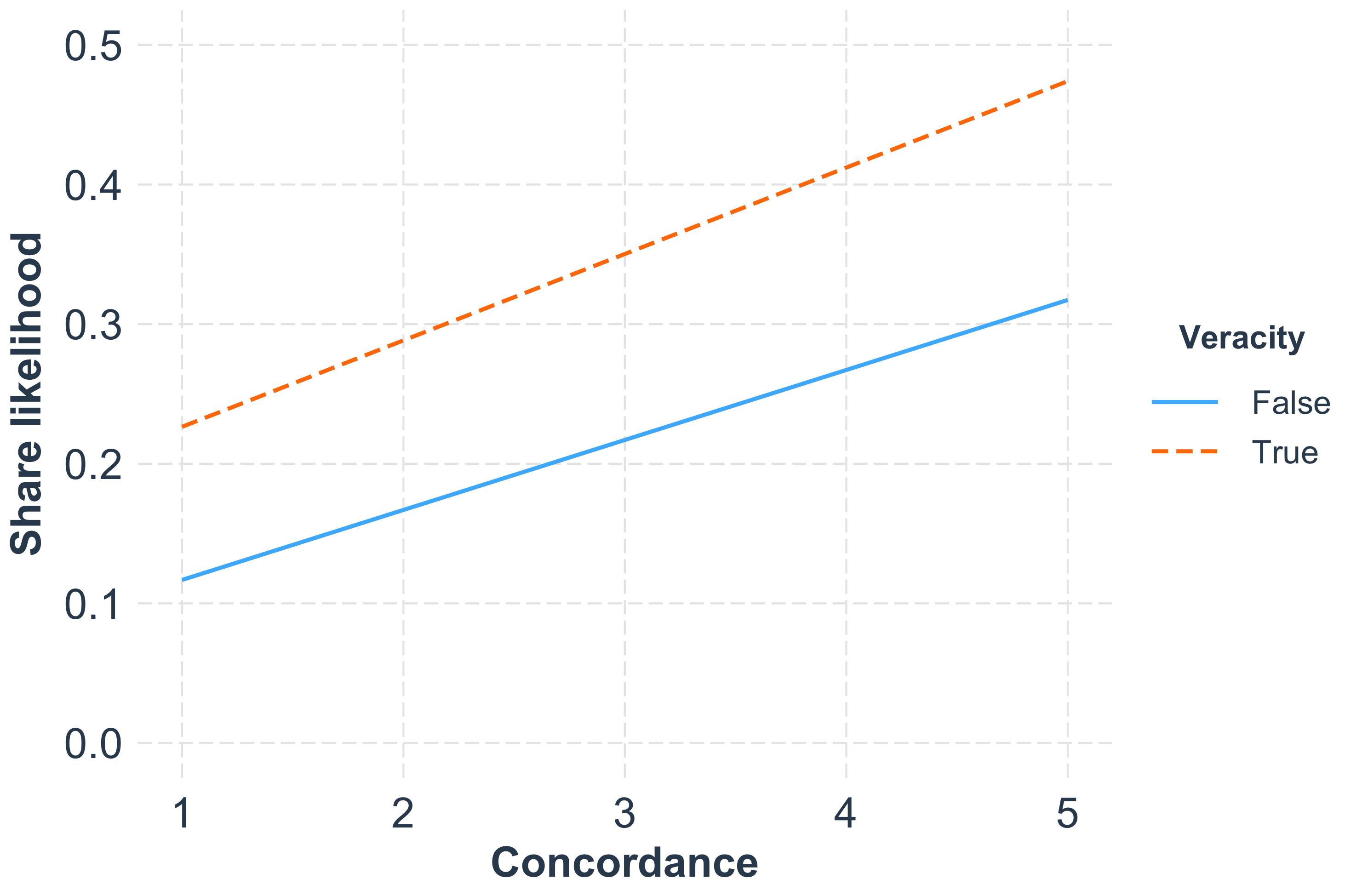}
 \captionof{figure}{The alignment of a headline's partisanship with the participant's increases the likelihood of sharing slightly more when the headline is true compared to when it is false.}
 \label{fig:concordance_by_veracity}
\end{minipage}\qquad
\begin{minipage}{.49\textwidth}
 \centering
 \includegraphics[width=\linewidth]{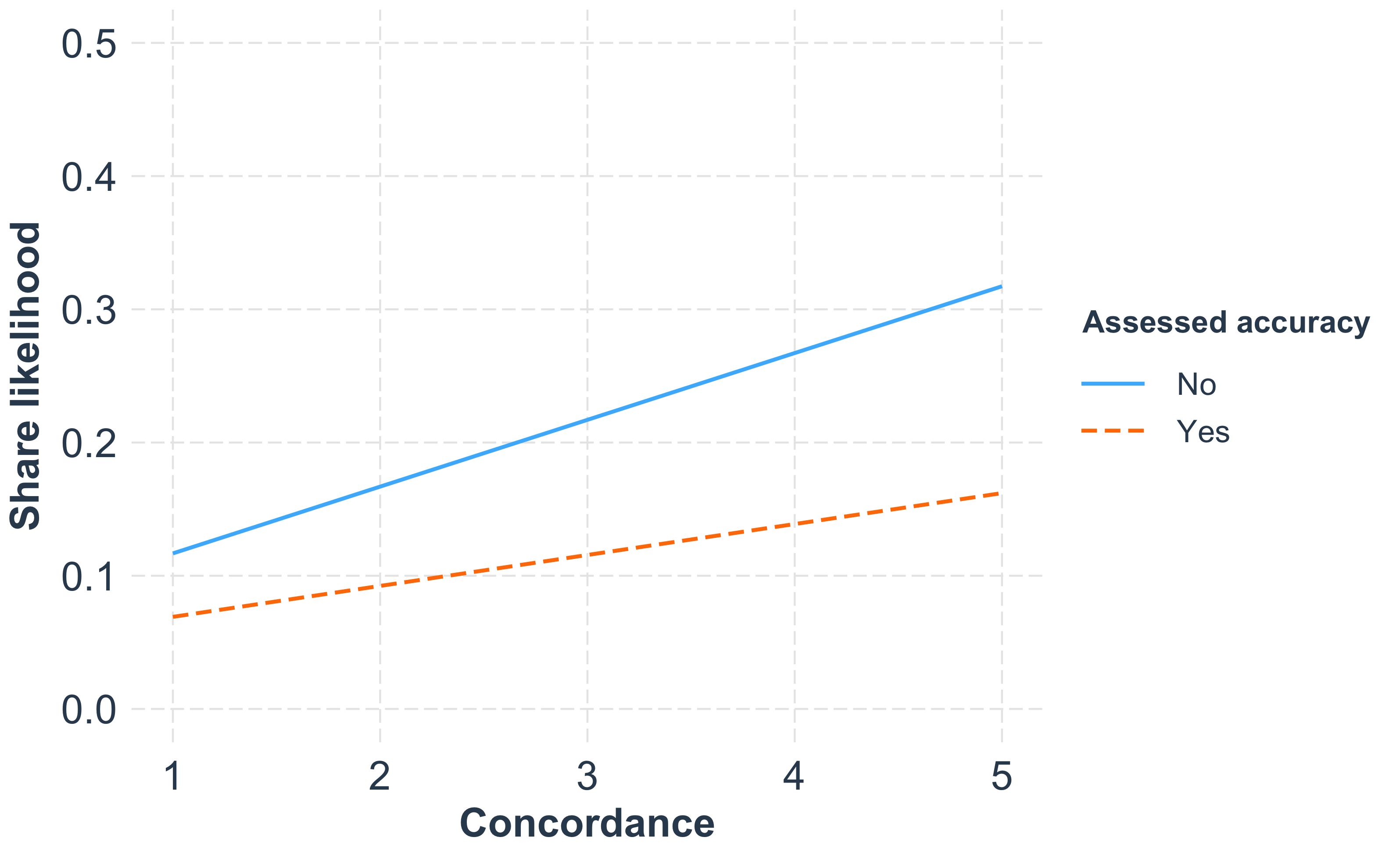}
 \captionof{figure}{Asking users to assess the accuracy of headlines restrains their sharing of headlines that are well-aligned with their partisanship.}
 \label{fig:concordance_by_accuracy_cond}
\end{minipage}
\end{figure}

The interaction between concordance and veracity was also significant [$\chi^2(1)=20.67$, $p<0.001$], indicating that the slope of share by concordance is different for false and true headlines. Figure~\ref{fig:concordance_by_veracity} indicates that the slope is slightly higher for true headlines, suggesting that the alignment of a headline's partisanship with the participant's increases sharing likelihood more when the headline is true compared to when it is false. Furthermore, we observed that the interaction between concordance and whether the participant was asked to assess accuracy was significant as well [$\chi^2(1)=8.23$, $p<0.01$]. As shown in Figure~\ref{fig:concordance_by_accuracy_cond}, asking users to assess the accuracy of headlines restrains their sharing of headlines that are well-aligned with their partisanship.

\begin{figure}[!t]
\centering
\begin{minipage}{.42\textwidth}
 \centering
 \includegraphics[width=\linewidth]{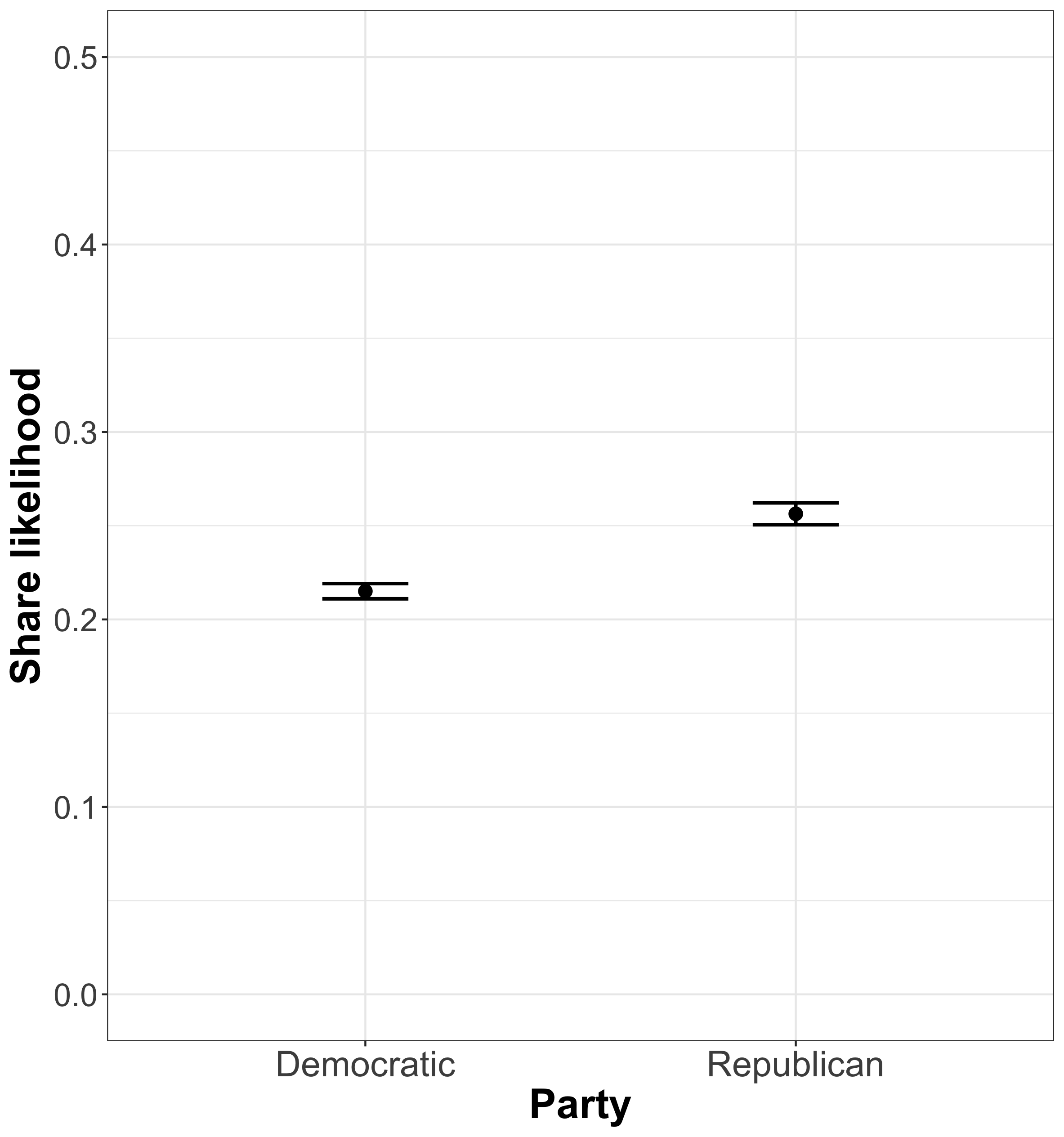}
 \captionof{figure}{Sample means of share likelihood by party. Republican participants were more likely to share headlines.}
 \label{fig:share_likelihood_by_party}
\end{minipage}\qquad
\begin{minipage}{.47\textwidth}
 \centering
  \includegraphics[width=\linewidth]{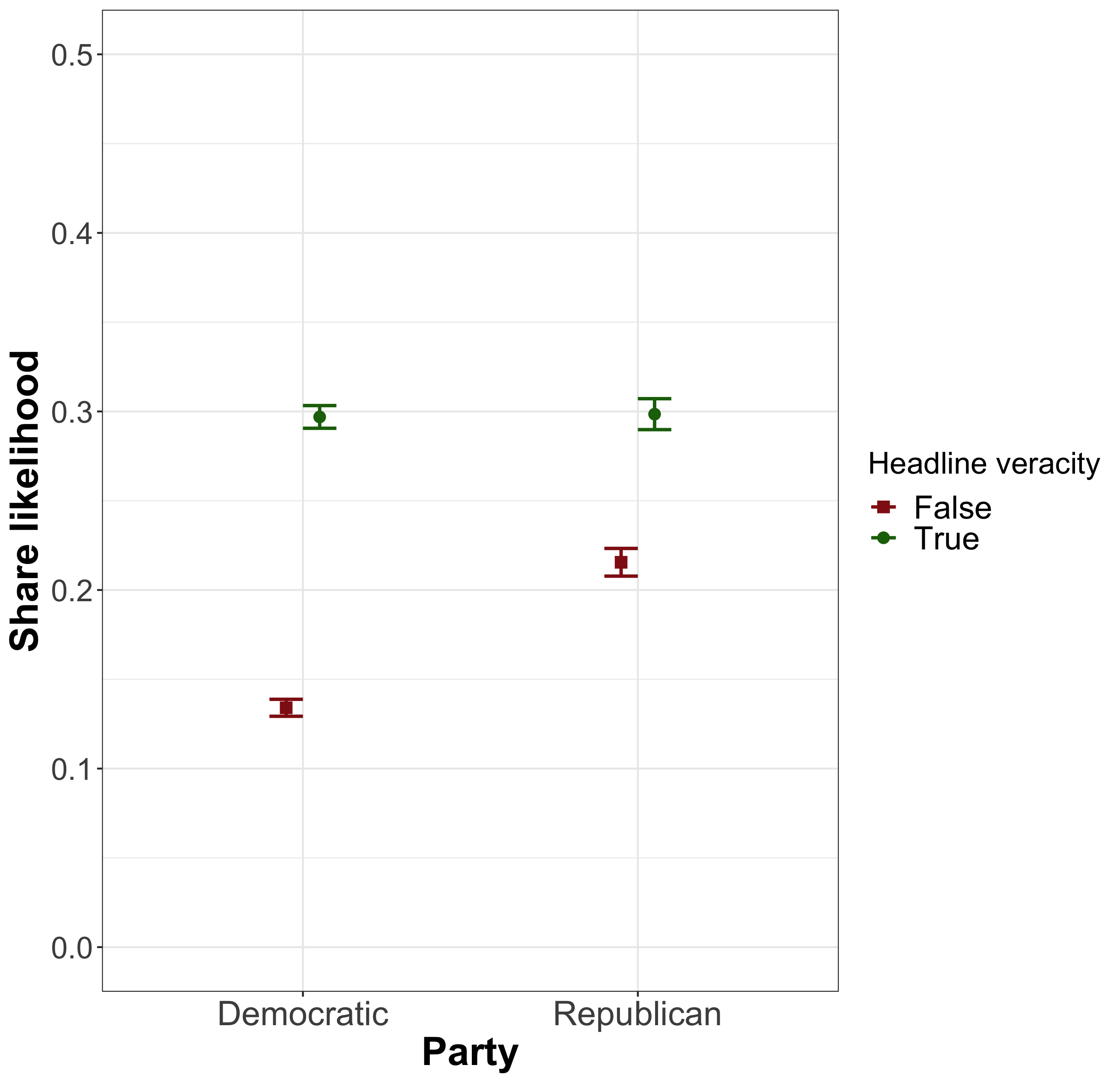}
 \captionof{figure}{Sample means of share likelihood by party and headline veracity. Democratic participants were less likely to share false headlines compared to Republicans.}
 \label{fig:party_veracity_interaction}
\end{minipage}
\end{figure}

\paragraph{Party} had a significant effect on sharing intentions [$\chi^2(1)=13.24$, $p<0.001$]. Figure~\ref{fig:share_likelihood_by_party} shows sample means of share likelihood by party. The figure suggests that Republicans share headlines more often than Democrats, irrespective of veracity. However, the interaction between party and headline veracity also had a significant effect [$\chi^2(1)=18.16$, $p<0.001$], with the means displayed in Figure~\ref{fig:party_veracity_interaction}. The figure indicates that while Democratic and Republican participants shared true headlines at a similar rate, Democratic participants were less likely to share false headlines compared to Republicans. This observations aligns with prior work that among other demographic factors investigated the association between Facebook users' party identification and the number of fake news stories they had shared \cite{grinberg2019fake}.

\begin{figure}[!t]
\centering
\begin{minipage}{.42\textwidth}
 \centering
 \includegraphics[width=\linewidth]{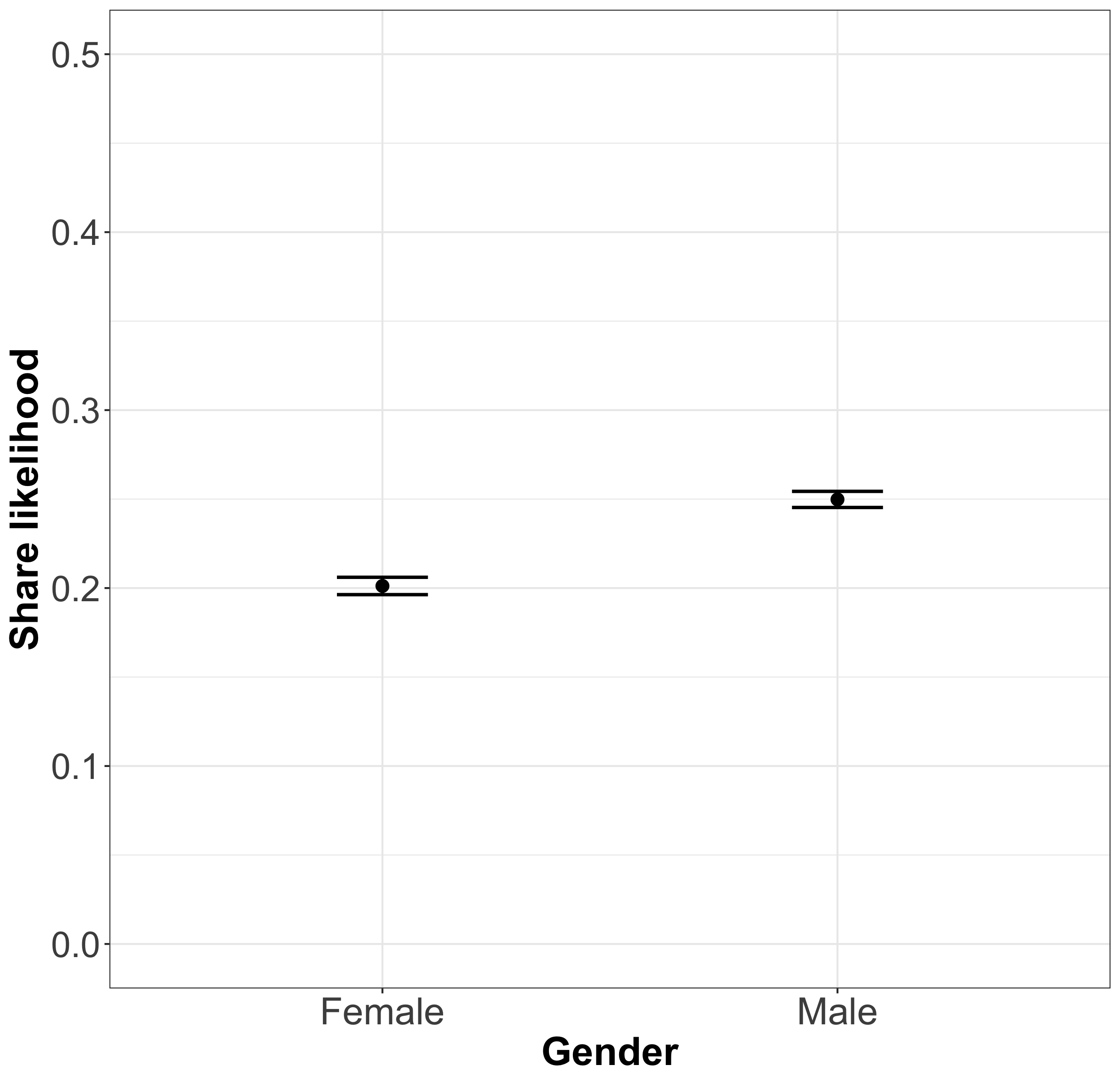}
 \captionof{figure}{Sample means of share likelihood by gender. Males are slightly more likely to share headlines compared to females.}
 \label{fig:share_likelihood_by_gender}
\end{minipage}\qquad
\begin{minipage}{.47\textwidth}
 \centering
 \includegraphics[width=\linewidth]{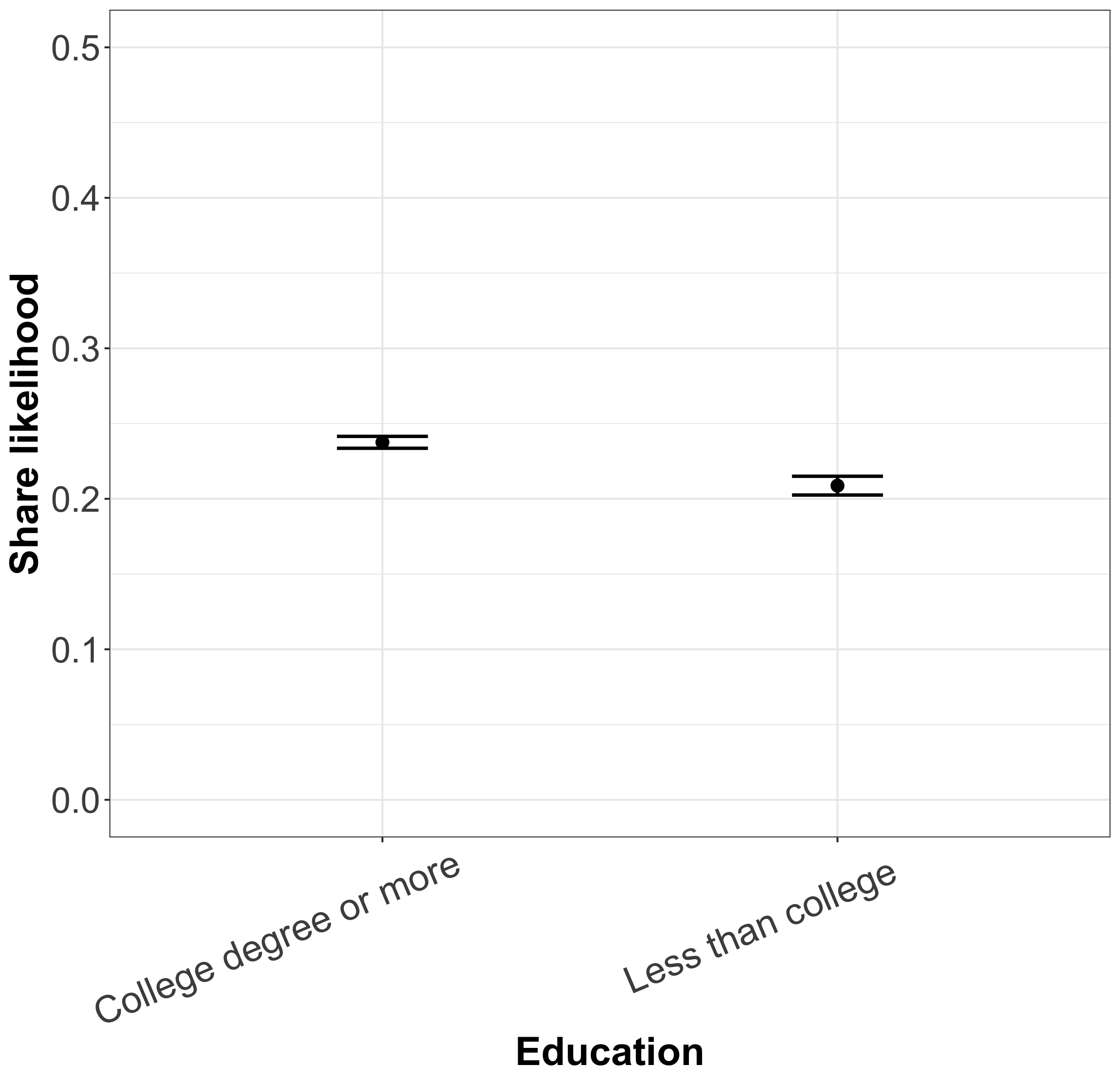}
 \captionof{figure}{Sample means of share likelihood by education. Participants who held a college degree were more likely to share headlines compared. The difference in share likelihood however, is small.}
 \label{fig:share_likelihood_by_education}
\end{minipage}
\end{figure}

\paragraph{Gender} had a significant effect on sharing intentions [$\chi^2(1)=11.23$, $p<0.001$], with the sample means shown in Figure~\ref{fig:share_likelihood_by_gender}. The figure suggests that males are more likely to share headlines compared to females.

\paragraph{Education} had a significant effect on likelihood of sharing at $\alpha=0.05$ level [$\chi^2(1)=5.85$, $p=0.02$]. As shown in Figure~\ref{fig:share_likelihood_by_education}, participants who held an Associate's degree or higher were more likely to share headlines compared to those that did not have a college degree.

\section{Cumulative Link Models}
\label{section:cumulative_link_model}

We tested the effects of our interventions on share intentions using the same formula as outlined in Section~\ref{section:nudge_main_results}, but using cumulative link mixed models instead of linear models. Cumulative link models are appropriate for fitting ordinal values and find the cumulative probability of the \textit{i}th rating (datapoint) falling in the \textit{j}th category or below. The categories in our data are ordered share decisions \textit{``No''}, \textit{``Maybe''}, and \textit{``yes''}. The cumulative link model assumes that there is a continuous but unobservable variable $Y_i$ with a mean that depends on the predictors and that this underlying distribution has a set of cut-points $\theta_1$, $\theta_2$, ..., $\theta_j$ where if $\theta_{k} < Y_i < \theta_{k+1}$, the manifest response (share decision) will take the value $k$.

Similar to the linear models from~\ref{section:nudge_main_results}, we developed a veracity model in which the independent variables were the main effects of the objective veracity of the headlines and our treatments (asking about accuracy, asking about reasoning, presenting checkboxes vs free-text to capture reasons), as well as the interaction between veracity and the treatments. In addition, we developed the cumulative link counterpart to the perceived accuracy model in~\ref{section:nudge_main_results}, which was fit to the data from the treatment conditions. In this model, the independent variables were participant's assessment of the accuracy of the headline, whether participants were asked to provide reasoning and whether they were presented with checkboxes, as well as the interaction between these treatments and accuracy assessment. In both models we included participant and claim as random effects.

To fit these models, we used the function ``clmm'' with a ``logit'' link from the package ``ordinal'' in R and set the threshold as symmetric. We then performed Likelihood Ratio Chi-Square tests (function ``Anova'' from package ``RVAideMemoire'') on each of the fitted models to determine whether the effects of the independent variables were significant. If we determined a factor was significant, we then performed a post-hoc Estimated Marginal Means (EMMeans) test across the levels of the factor of interest averaging over all other factors. We used the function ``emmeans'' from the R package ``emmeans'' with mode ``mean.class'' to obtain and compare the expected values of the ordinal response on a scale of 1 to 3 (the number of categories) for each of the levels of the factor of interest. P values were adjusted with Tukey method to account for multiple comparisons. The results of the cumulative link models were consistent with the results obtained from the linear models in~\ref{section:nudge_main_results}.

Similar to the results we observed for the linear model counterparts, the effects of veracity in the veracity model and perceived accuracy in the perceived accuracy model were both significant [$\chi^2(1)=29.80$, $p<0.001$ for veracity model, $\chi^2(1)=1851.56$, $p<0.001$ for perceived accuracy model]. The EMMeans showed that participants were more likely to have the intention of sharing objectively true rather than false headlines [$z=5.03$, $p<0.001$, $E(False)=1.23$, $E(True)=1.48$]. Similarly, they were more likely to share headlines that they perceived as true [$z=13.49$, $p<0.001$, $E(\textit{Perceived as false})=1.05$, $E(\textit{Perceived as true}=1.58$].

Similarly, providing accuracy assessments had a significant effect on participants' likelihood of sharing either false or true headlines [$\chi^2(1)=33.83$, $p<0.001$]. The EMMeans test revealed that participants were more likely to share headlines if they were not asked about their accuracy [$z=4.89$, $p<0.001$, $E(\textit{Accuracy not provided})=1.44$, $E(\textit{Accuracy provided})=1.28$].

In addition, the effects of providing reasoning and the format of reasoning were significant in both the veracity and the perceived accuracy models [providing reasoning: $\chi^2(1)=13.05$, $p<0.001$ for veracity, $\chi^2(1)=13.38$, $p<0.001$ for perceived accuracy; reasoning format: $\chi^2(1)=6.21$, $p=0.01$ for veracity, $\chi^2(1)=5.14$, $p=0.02$ for perceived accuracy]. Participants were more likely to share headlines if they were not asked to provide their reasoning about why the claim was or was not accurate [$z=3.52$, $p<0.001$, $E(\textit{Reasoning not provided})=1.41$, $E(\textit{Reasoning provided})=1.30$ for veracity; $z=3.38$, $p<0.001$, $E(\textit{Reasoning not provided})=1.36$, $E(\textit{Reasoning provided})=1.26$ for perceived accuracy]. Providing reasons via the checkbox set of reason categories also lowered their likelihood of sharing content [$z=2.60$, $p=0.01$, $E(\textit{Free-text})=1.40$, $E(\textit{Checkbox})=1.32$ for veracity; $z=2.58$, $p=0.01$, $E(\textit{Free-text})=1.35$, $E(\textit{Checkbox})=1.27$ for perceived accuracy].

The veracity model also indicated that the interaction between veracity and providing accuracy is statistically significant [$\chi^2(1)=10.98$, $p<0.001$]. The interaction however, was not practically meaningful [$E(\textit{False, Accuracy not provided})=1.32$, $E(\textit{False, Accuracy provided})=1.15$, $E(\textit{True, Accuracy not provided})=1.57$, $E(\textit{True, accuracy provided})=1.40$].

\section{Investigation of Potential Confounds in the Nudge Study}
\label{section_potential_confounds}

\subsection{Makeup of Data and Impact of Removing Spams.}
The task in the Nudge study presented 10 claims to each participant but because some participants abandoned the task before its conclusion, we had fewer data from them. It is conceivable that if the attrition rate is different across conditions, then the conditions differ not only in what treatment they received, but also in what type of people contributed more data to each condition. 
Therefore, we probed how many participants per condition did not finish all the 10 headlines. This number across all conditions was in the range of 30-50, suggesting that the dropout rate was similar. We then analyzed the spam rate across conditions which was more variant (condition 1: 145 , condition 2: 114 , condition 3: 136, condition 4: 174). It is possible that different interventions result in different spam rates and that those participants who stay and work through a more laborious condition, are in fact, characteristically different from those who finish the task by spamming.

We performed a Pearson’s Chi Square test to investigate if the distribution of spams was different from a uniform distribution. The difference was statistically significant [$\chi^2(3)=13.02$, $p=0.005$], suggesting that the conditions may have had a role in different numbers of users becoming spammers across conditions.
We then analyzed the share rate in spams across different conditions which was similar (see Table ~\ref{tab:spam_means}), with share mean of approximately 0.72 across all conditions regardless of headline veracity. 

In the main section of the paper, we have presented our findings above excluding the Spams. However, we perform the same analyses including the spam datapoints in the Appendix section~\ref{section:spams}. Some of our results that pertain to conditions that have heavier interventions are no longer statistically significant when spams are included. The reason is that in these conditions, the share rates are low and therefore the difference between conditions is smaller but detectable in the absence of noise. Including noise, i.e., spams, in a condition, increases the sample size while adding a relatively large number of positive datapoints, or datapoints that indicate a positive intention of sharing, for both true and false headlines. The difference that existed before will now be diluted.

\subsection{Deliberation Priming.}
Although in the control condition for the Taxonomy study we did not have any of the accuracy and reasoning nudges, after each sharing decision, we asked participants why they would or would not share the article. This question by itself may have acted as a deliberation prime on subsequent sharing decisions. To test this hypothesis, we developed a model with share intention as the dependent variable and veracity and whether the item was the first item presented to the user as independent variables and included participant identifier as a random effect. We fit the model to the first and last datapoints that participants in the control condition provided. We found that as expected, veracity was positively correlated with sharing intentions and the correlation was statistically significant [$\beta=0.16, p<0.001$]. Being the first decision by the participant also had a positive albeit nonsignificant correlation with sharing [$\beta=0.49, p=0.23$]. The interaction between the two had a negative and nonsignificant correlation [$\beta=-0.08, p=0.20$]. Despite the lack of significance, we observe that the effect of veracity on the last item presented to the user is twice as large as that of the first item [$0.16$ for the last decision, $0.16-0.08=0.08$ for the first]. This observation gives some degree of support to the hypothesis that simply asking users to ponder over their sharing decision may have primed them to be mindful of the headline's accuracy over time.

\subsection{Investigating Potential Learning Effects of Repeated Accuracy Assessments}

We wished to investigate whether making repeated judgements about accuracy had a learning effect on participants leading their subsequent accuracy assessments to be closer to the headline's actual veracity. Therefore, we fit the following model to the first and last datapoints that participants in the treatment conditions had provided:
\begin{multline}
    \text{accuracy assessment} == \text{veracity} \sim  \text{is first question} \times ( \text{reasoning condition} + \text{reasoning format})\\ + (1|\text{participant}) + (1|\text{claim})
\end{multline}

Because the outcome of the model was dichotomous (1 if veracity and perceived accuracy matched, 0 if they did not), we used the function ``glmer'' with link ``logit'' from the R package ``lme4'' to fit the data. A Wald Chi-Square test on the model revealed that the effect of whether the participant's judgement was the first or the last was in fact not significant [$\chi^2(1)=0.07$, $p=0.80$], indicating that repeated judgements had not been a significant confounder in the study.

\section{Spams}
\label{section:spams}
In this section, we include the spam datapoints of the Nudge study in the dataset and perform the same analyses that we conducted in the Results section.

The Wald-Chi Square tests fitted to our linear models revealed that the effect of veracity in the veracity model and the effect of perceived accuracy in the perceived accuracy model on sharing intention were both significant [$\chi^2(1)=30.13$, $p<0.001$ for veracity, $\chi^2(1)=2645.61$, $p<0.001$ for perceived accuracy]. Post-hoc Estimated Marginal Means tests revealed that participants were more likely to share an objectively true headline compared to a false one [$z=4.60, p<0.001$]. Similarly, they were more likely to share a headline that they perceived as true rather than one they perceived as false [$z=42.04, p<0.001$]

\subsection{Effect of Providing Accuracy Assessments}
Consistent with the results we observed when excluding spams, providing accuracy assessment had a significant effect on sharing intentions for the veracity model [$\chi^2(1)=32.04$, $p<0.001$].

Figure ~\ref{fig:all_3_w_spams} shows how sharing rates differ in conditions 1 and 2 by whether participants were asked about accuracy and headline veracity. Asking people to provide accuracy assessments decreases sharing of false headlines by 29\% while the reduction in sharing of true content is 17\%. 

\subsection{Effect of Providing Reasoning}

The effect of reasoning on sharing intention when including spams was not significant in either the veracity or the perceived accuracy models [$\chi^2(1)=0.40$, $p=0.53$ for the veracity model, $\chi^2(1)=0.50$, $p=0.48$ for the perceived accuracy model]. Similarly, the interaction effect of reasoning and veracity was not significant in the veracity model [$\chi^2(1)=1.13$, $p=0.29$]. However, the effect of interaction between reasoning and perceived accuracy was significant [$\chi^2(1)=7.25$, $p=0.007$].

Figure ~\ref{fig:all_3_w_spams} shows sharing rate means across true and false headlines for conditions 2 and 3 which differ in whether participants provided reasoning. Figure~\ref{fig:subjective_spams} shows that the sharing of headlines that are perceived as true is reduced when reasoning is requested. The means in sharing rate of headlines perceived as false however, do not vary much across the 2 conditions.

\subsection{Effect of Reasoning Format}

We observed that the effect of reasoning form when including spams was not significant in either the veracity or the perceived accuracy models [$\chi^2(1)=0.61$, $p=0.43$ in the veracity model, $\chi^2(1)=0.52$, $p=0.47$ in the perceived accuracy model]. The effect of the interaction between veracity and reasoning form was not significant either [$\chi^2(1)=2.92$, $p=0.09$]. However, the interaction between reasoning form and perceived accuracy was significant [$\chi^2(1)=14.81$, $p<0.001$].

\begin{figure}[t]
 \centering
 \includegraphics[width=0.8\linewidth]{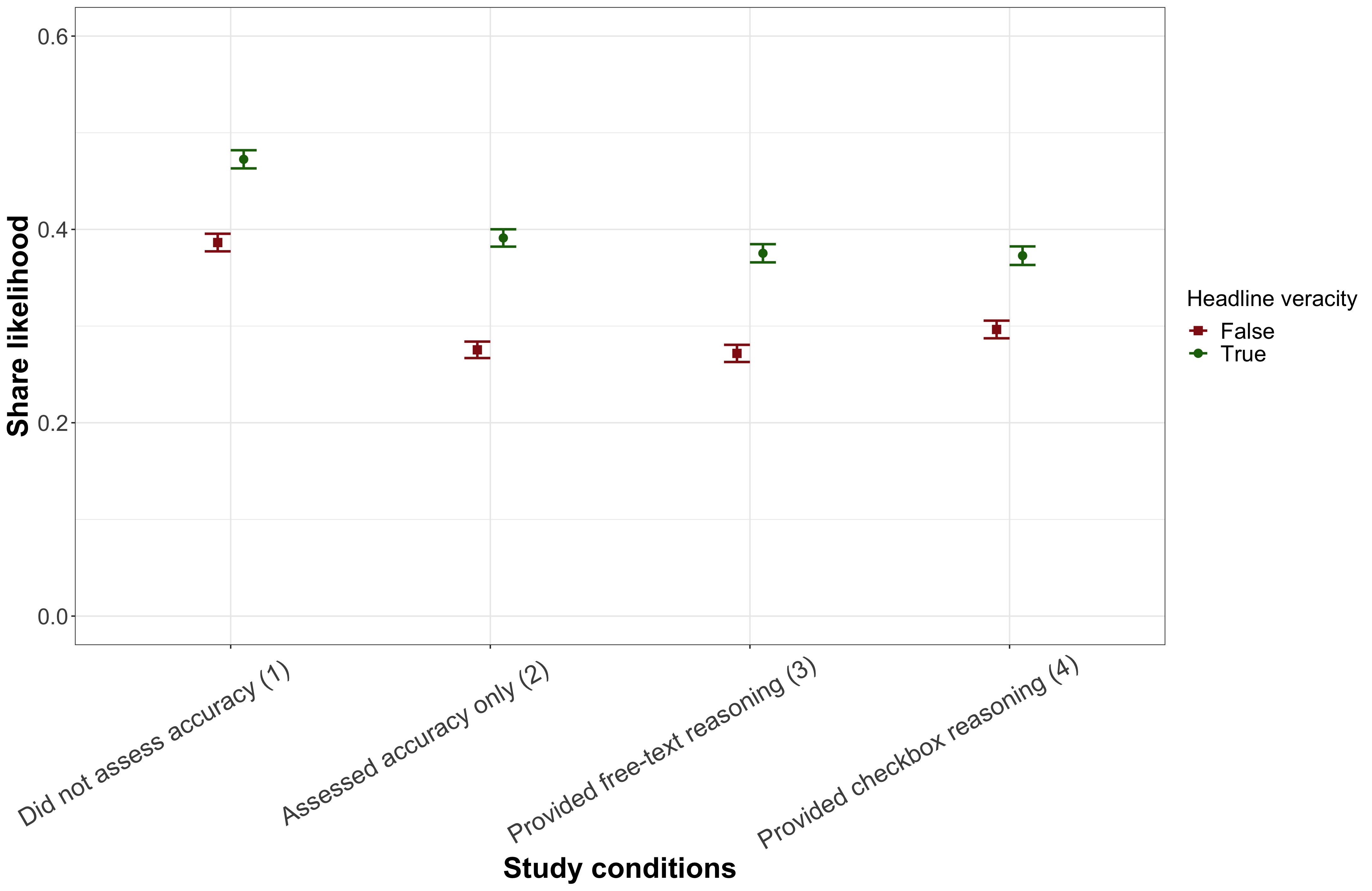}
 \caption{Share rate of true and false headlines across study conditions including spam datapoints. The results suggest that people are less likely to share both accurate and inaccurate content if they are asked to assess the content's accuracy although the reduction in shared false content is higher (condition 1 vs 2). However, asking people to provide their reasoning in addition to assessing accuracy does not result in a statistically significant difference compared to if they only assess accuracy (condition 2 vs 3). Similarly, there does not exist a statistically significant difference in means of sharing true and false content across the reasoning format conditions (condition 3 vs 4).}
 \Description{}
 \label{fig:all_3_w_spams}
\end{figure}

\begin{figure}[t]
 \centering
 \includegraphics[width=0.7\linewidth]{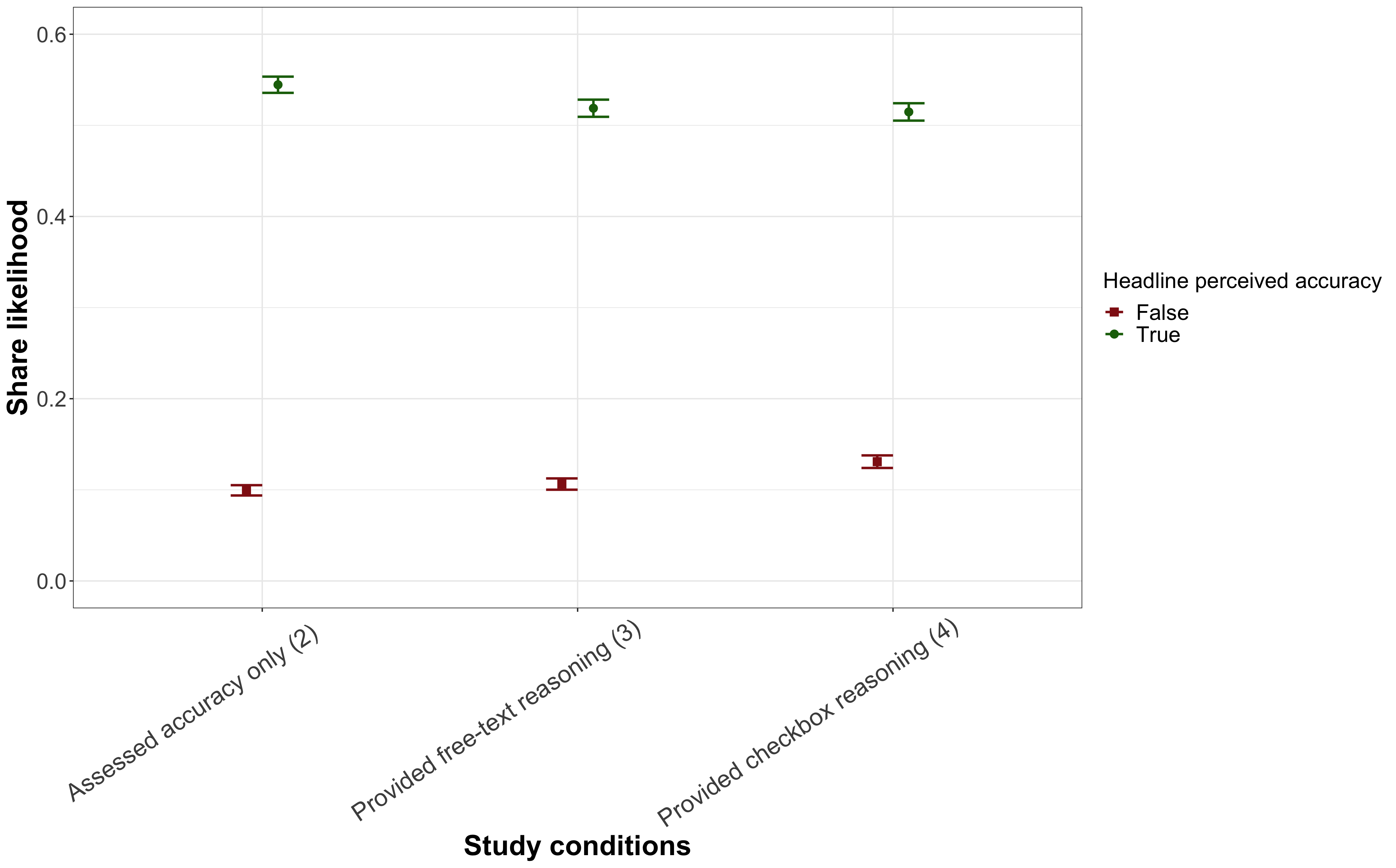}
 \caption{Share rate of headlines across study conditions including spam datapoints for headlines that were perceived as true or false. In condition 3 where participants where asked about their rationales, the share mean for headlines perceived as true was decreased compared to condition 2 (condition 2 vs 3). However, asking people people about their rationales via a checkbox increases sharing of content that they initially perceived as false (condition 3 vs 4).}
 \Description{}
 \label{fig:subjective_spams}
\end{figure}

Figure~\ref{fig:all_3_w_spams} shows the means across the conditions with different instruments for capturing reasoning. Figure~\ref{fig:subjective_spams} shows that in condition 4 where checkboxes were presented, people shared headlines that they initially perceived as false at a higher rate compared to condition 3.

\begin{table*}[!t]
\setlength{\aboverulesep}{0pt}
\setlength{\belowrulesep}{0pt}
\setlength{\extrarowheight}{.35ex}
\newcolumntype{g}{>{\columncolor{gray!16}}c}
\setlength{\tabcolsep}{6pt}
\caption{Share means in spam entries across experimental conditions and headline veracity.}
\label{tab:spam_means}
\centering
\begin{tabular}{gcccc}
 \toprule
 \rowcolor{gray!16} Veracity. & {Cond. 1} & {Cond. 2} & {Cond. 3} & {Cond. 4} \\
 \midrule
 True & 0.76 & 0.70 & 0.67 & 0.70\\
 False & 0.74 & 0.72 & 0.71 & 0.73\\
 \bottomrule
\end{tabular}
\end{table*}%

\section{Headlines Used in the Study of Behavioral Nudges}
\label{section:headlines_table_nudge_study}

We present the headlines that we used in the user study of behavioral nudges along with their veracity and partisanship. Partisanship of a headline was rated by each participant that was presented the headline in the study. The partisanship measure in the table is an average over all these ratings on scale of -2 (more favorable for Democrats) to 2 (more favorable for Republicans).

\afterpage{
\begin{footnotesize}
\begin{longtable}{c|p{8.8cm}|c|c}
\caption{The headlines used in the user study of behavioral nudges. Partisanship is on a scale of -2 (more favorable for Democrats) to 2 (more favorable for Republicans) }
 \label{tab:taxonomy}
 \\

 \hline \rowcolor{gray!28}
 \multicolumn{1}{l|}{\textbf{Topic}} & \multicolumn{1}{l|}{\textbf{Headline}} & 
 \multicolumn{1}{l|}{\textbf{Veracity}} & \multicolumn{1}{l|}{\textbf{Partisanship}} \\ \hline 
 \endfirsthead
 
 \multicolumn{4}{c}%
 {{\tablename\ \thetable{} -- continued from previous page}} \\
 \hline \rowcolor{gray!28} 
 \multicolumn{1}{l|}{\textbf{Category}} & 
 \multicolumn{1}{l|}{\textbf{Headline}} & 
 \multicolumn{1}{l|}{\textbf{Veracity}} & \multicolumn{1}{l|}{\textbf{Partisanship}} \\ \hline 
 \endhead
 
 \hline \multicolumn{3}{|r|}{{Continued on next page}} \\ \hline
 \endfoot
 
 \hline \hline
 \endlastfoot

   \multirow{29}{*}{\rotatebox[origin=c]{90}{\centering Politics}} &
  
 
    Asylum-Seekers Can Apply at U.S. Embassies Abroad & False & $0.70$ \\
    \cline{2-4}
    & Migrants Are `Free to Leave Detention Centers Any Time' & False & $0.54$ \\
    \cline{2-4}
    & Joe Biden Said Poor Kids Are `Just as Talented as White Kids' & True & $0.68$\\
    \cline{2-4}
    & Sanders Proposed Raising Taxes to 52\% on Incomes Over \$29,000  & False & $1.10$ \\
    \cline{2-4}
    & Trump Proposed Cuts to Federal Pay Raises, Citing ``Serious Economic Conditions'' & True & $-0.06$ \\
    \cline{2-4}
    & Obama Admin Built Cages That House Immigrant Children at U.S.-Mexico Border & True & $1.25$\\
    \cline{2-4}
    & Sarah Palin Said ``US Already Attacked Iran Back When It Called Itself Iraq'' & False & $-0.55$ \\
    \cline{2-4}
    & U.S. Sen. Kamala Harris Said ``White Lab Coats Are a Sign of Doctors' Racism'' & False & $0.68$\\
    \cline{2-4}
    & Seniors on Social Security Have to Pay for Medicare While ``Illegal Immigrants'' Get It Free & False & $1.01$ \\
    \cline{2-4}
    & Gun Violence Killed More People in U.S. in 9 Weeks than U.S. Combatants Died in D-Day & True & $-1.13$\\
    \cline{2-4}
    & Border Wall Construction Threatened Native American Burial Sites & True & $-1.12$\\
    \cline{2-4}
    & 56\% of Survey Respondents Said `Arabic Numerals' Shouldn't be Taught in School & True & $0.06$\\
    \cline{2-4}
    & Biden's Campaign Demanded an American Flag Be Removed from a Library & False & $1.21$\\
     \cline{2-4}
    & The Obamacare Website Cost \$5 Billion & False & $1.36$\\
     \cline{2-4}
    & ABC, CBS, and NBC Blacked Out Pam Bondi's Legal Defense of Trump during His Impeachment Trial & False & $0.65$\\
    \cline{2-4}
    & President Obama Ordered More Than 500 Drone Strikes & True & $0.91$\\
    \cline{2-4}
    & Eric Trump Tweeted About Iran Strike Before It Was Made Public & False & $-1.03$\\
    \cline{2-4}
    & the NRA Opposed Reauthorization of the Violence Against Women Act in April 2019 & True & $-0.78$\\
    \cline{2-4}
    & 
    10,150 Americans Were Killed by Illegal Immigrants in 2018 & False & $1.33$\\
    \cline{2-4}
    & President Trump's Awarding of a Purple Heart to a Wounded Vet Went Unreported by News Media & True & $1.24$\\
    \cline{2-4}
    &  The New Way Forward Act Would Protect Criminals from Deportation & True & $0.56$\\
    \cline{2-4}
    & US Intelligence Eliminated a Requirement That Whistleblowers Provide Firsthand Knowledge & False & $0.22$\\
    \cline{2-4}
    & New York Reprimanded Trump Family for `Stealing from a Children's Cancer Charity' & False & $-1.59$\\
    \cline{2-4}
    & Age Matters More than Sexual Orientation to U.S. Presidential Voters, Poll Finds & True & $-0.20$ \\
    \cline{2-4}
    & Watchdog: ICE Doesn't Know How Many Veterans It Has Deported & True & $-1.01$ \\
    \cline{2-4}
    & Experts and Officials Warned in 2018 US Couldn't Respond Effectively to a Pandemic & True & $-1.23$\\
    \cline{2-4}
    & Trump Said These About the COVID-19 Pandemic in the Span of Eight Days \break [Presented with a picture of two quotes] & True & $-0.98$\\
    \cline{2-4}
    & The CDC Significantly Readjusted COVID-19 Death Numbers Down & False & $0.29$\\
\cline{1-4}    

   \multirow{10}{*}{\rotatebox[origin=c]{90}{Science \& Tech}} &


 
 A Scientist Was Jailed After Discovering a Deadly Virus Delivered Through Vaccines & False & $0.25$\\
    \cline{2-4}
    & Rain That Falls in Smoky Areas After a Wildfire Is Likely to Be ``Extremely Toxic'' & False & $-0.30$\\
    \cline{2-4}
    & A Scientific Study Proved That ``Conspiracists'' Are ``The Most Sane of All'' & False & $0.24$\\
    \cline{2-4}
    & Scientists Were Caught Tampering with Raw Data to Exaggerate Sea Level Rise & False & $0.98$ \\
    \cline{2-4}
    & Women Retain DNA From Every Man They Have Ever Slept With & False & $0.21$\\
    \cline{2-4}
    & Chinese Doctors Confirm African People Are Genetically Resistant to Coronavirus & False & $0.17$\\
    \cline{2-4}
    & Abortions are Linked to an Increased Risk of Breast Cancer & False & $1.00$\\
    \cline{2-4}
    & Global Sea Ice is at a Record-Breaking Low & True & $-0.86$\\
    \cline{2-4}
    & A 17-Year-Old Eagle Scout Built a Nuclear Reactor in His Mom's Backyard & True & $0.04$\\
    \cline{2-4}
    & A Study Showed That Dogs Exhibit Jealousy & True & $-0.01$\\
    \cline{2-4}

\multirow{11}{*}{\rotatebox[origin=c]{90}{}} &
 No Two Snowflakes Are Exactly Alike & True & $0.01$\\
\cline{2-4}
    & These Photographs Show the Same Spot in the Arctic 100 Years Apart & True & $-1.01$\\
\cline{2-4}
    & There Is a Point in the Ocean Where the Closest Human Could Be an Astronaut & True & $0.00$ \\
\cline{2-4}
    & There Are More Trees on Earth Than Stars in the Milky Way & True & $0.04$\\
\cline{2-4}
    & Tibetan Monks Can Raise Body Temperature With Their Minds & True & $-0.04$\\
\cline{2-4}
    & Presidential Alerts Give the Government Total Access to Your Phone & False & $-0.35$\\
    \cline{2-4}
    & A New Study Showed That Marijuana Leads to a Complete Remission of Crohn's Disease & False & $-0.44$\\
    \cline{2-4}
    & Marijuana Use Can Lead to Simultaneous Screaming and Vomiting & True & $0.48$\\
\cline{2-4}
    &  Some Phone Cameras Inadvertently Opened While Users Scrolled Facebook App  & True & $0.08$ \\
\cline{2-4}
    & Sipping Water Every 15 Minutes Will Prevent a Coronavirus Infection & False & $0.21$\\
    \cline{2-4}
    & Regular Consumption of Lemons and Hot Water Helps against the Spread of COVID-19 & False & $0.17$\\
  \hline

\end{longtable}
\end{footnotesize}
}

\received{June 2020}
\received[revised]{October 2020}
\received[accepted]{December 2020}

\end{document}


\maketitle

\section{Nudge Study Questionnaire}
The following is the questionnaire that was presented to the participants in the Nudge study. The first section (Sharing), was presented for each of the 10 claims participants were shown. Not all questions in the Sharing section were shown to all participants. If a participant was in any of the treatment conditions, for each claim, the survey would ask whether the claim was accurate or inaccurate and how confident the participant was in their belief. If the participant was in one of the reasoning conditions, the survey would additionally ask why they believed the claim was (in)accurate. The questions in the Partisanship section were shown for each claim that the participant had previously seen in the Sharing section.

\subsection{Sharing}
To the best of your knowledge, is the claim in the above headline accurate or inaccurate?
\vspace{-0.25cm}
\begin{itemize}
    \item Inaccurate
    \vspace{-0.25cm}
    \item Accurate
\end{itemize}

\noindent How confident are you in your belief?
\vspace{-0.25cm}
\begin{itemize}
    \item Not at all confident
    \vspace{-0.25cm}
    \item Not very confident
    \vspace{-0.25cm}
    \item Fairly confident
    \vspace{-0.25cm}
    \item Very confident
\end{itemize}

\noindent [One of the following questions was presented depending on in which condition the participant was placed and their answer to the accuracy question.]
\vspace{0.25cm}

\noindent Please explain why you believe the claim is accurate [or inaccurate].
\vspace{0.25cm}

\noindent Why do you think the headline is accurate? Select all that apply and explain your choice for each one. 
\vspace{-0.25cm}
\begin{itemize}
    \item Accurate on the evidence
    \vspace{-0.2cm}
    \begin{itemize}
        \item I have a high degree of knowledge on this topic that allows me to assess this claim myself (e.g., I teach/write about this topic or I use this in my work).
        \item I have firsthand knowledge on the subject or am an eyewitness.
        \item My other trusted sources (besides the source of this article) confirm the entire claim. (Please mention the sources)
        \item The claim is from a source I trust.
        \item Evidence presented in the article corroborates the claim.
    \end{itemize}
    
    \item Plausible
    \vspace{-0.2cm}
    \begin{itemize}
        \item The claim is consistent with my past experience and observations.
    \end{itemize}
    
    \item Don't know
    \vspace{-0.2cm}
    \begin{itemize}
        \item I'm not sure, but I want the claim to be true.
        \item I was just guessing.
    \end{itemize}
   
   \item Other
   
\end{itemize}

\noindent Why do you think the headline is inaccurate? Select all that apply and explain your choice for each one.
\vspace{-0.25cm}
\begin{itemize}
    \item Inaccurate by contrary knowledge
    \vspace{-0.2cm}
    \begin{itemize}
        \item I have a high degree of knowledge on this topic that allows me to assess this claim myself (e.g., I teach/write about this topic or I use this in my work).
        \item I have firsthand knowledge on the subject or am an eyewitness.
        \item The claim contradicts some information related to the case that I know from trusted sources.
    \end{itemize}
    
    \item Implausible
    \vspace{-0.2cm}
    \begin{itemize}
        \item The claim is not consistent with my past experience and observations.
        \item If this were true, I would have heard about it.
        \item The claim appears to be inaccurate based on its presentation (its language, flawed logic, etc.).
        \item The claim appears motivated or biased.
        \item The claim references something that is impossible to prove.
    \end{itemize}
    
    \item Don't know
    \vspace{-0.2cm}
    \begin{itemize}
        \item I'm not sure, but I do not want the claim to be true.
        \item I was just guessing.
    \end{itemize}
    
    \item Other
   
\end{itemize}

\noindent In addition to the above, does the claim have any of the following characteristics?
\vspace{-0.25cm}
\begin{itemize}
    \item The claim is misleading.
    \vspace{-0.25cm}
    \item The claim is not from a source I trust.
\end{itemize}

\noindent [The following questions were presented to all participants in all conditions.]
\vspace{0.25cm}

\noindent Would you consider sharing the article on social media (for example on Facebook or Twitter)?
\vspace{-0.25cm}
\begin{itemize}
    \item Yes
    \vspace{-0.25cm}
    \item No
    \vspace{-0.25cm}
    \item Maybe
\end{itemize}

\noindent Why would you consider sharing the article? [or wouldn't -- depending on what the participant selected as their answer to their previous question.]

\subsection{Trust Practices and Judgements of News Accuracy}
\noindent Suppose you saw a headline similar to the ones in this survey and you were unsure about its accuracy. If you were curious about the accuracy, how would you go about verifying the claim?
\vspace{0.25cm}

\noindent How comfortable or uncomfortable would you be sharing your judgment on the accuracy of the claims public to your friends?
\vspace{-0.25cm}
\begin{itemize}
    \item Very uncomfortable
    \vspace{-0.25cm}
    \item Somewhat uncomfortable
    \vspace{-0.25cm}
    \item Somewhat comfortable
    \vspace{-0.25cm}
    \item Very comfortable
\end{itemize}

\noindent How comfortable or uncomfortable would you be sharing your judgment on the accuracy of the claims public to the world?
\vspace{-0.25cm}
\begin{itemize}
    \item Very uncomfortable
    \vspace{-0.25cm}
    \item Somewhat uncomfortable
    \vspace{-0.25cm}
    \item Somewhat comfortable
    \vspace{-0.25cm}
    \item Very comfortable
\end{itemize}

\noindent Think of some news publishing media, public figures, friends, and relatives that you trust. Tell us why you trust them.
\vspace{0.25cm}

\noindent Think of some news publishing media, public figures, friends, and relatives that you *don't* trust. Tell us why you don't trust them.
\vspace{0.25cm}

\noindent Did you respond randomly at any point while completing the news assessment task? 
Note: Please be honest! You will get your HIT regardless of your response.
\vspace{-0.25cm}
\begin{itemize}
    \item Yes
    \vspace{-0.25cm}
    \item No
\end{itemize}

\noindent Did you search the internet (via Google or otherwise) for any of the headlines?
Note: Please be honest! You will get your HIT regardless of your response.
\vspace{-0.25cm}
\begin{itemize}
    \item Yes
    \vspace{-0.25cm}
    \item No
\end{itemize}

\subsection{Partisanship}
Assuming the above headline is entirely accurate, how favorable would it be to Democrats versus Republicans?
\vspace{-0.25cm}
\begin{itemize}
    \item More favorable for Democrats
    \vspace{-0.25cm}
    \item Somewhat more favorable for Democrats
    \vspace{-0.25cm}
    \item Equally favorable for Democrats and Republicans
    \vspace{-0.25cm}
    \item Somewhat more favorable for Republicans
    \vspace{-0.25cm}
    \item More favorable for Republicans
\end{itemize}

\subsection{Cognitive Reflection Test}
\noindent In the next section, you will be asked a set of 7 word problems. Please do your best to answer as accurately as possible. 
\vspace{0.25cm}

\noindent The ages of Mark and Adam add up to 28 years total. Mark is 20 years older than Adam. How many years old is Adam?
\vspace{0.25cm}

\noindent If it takes 10 seconds for 10 printers to print out 10 pages of paper, how many seconds will it take 50 printers to print out 50 pages of paper?
\vspace{0.25cm}

\noindent On a loaf of bread, there is a patch of mold. Every day, the patch doubles in size. If it takes 40 days for the patch to cover the entire loaf of bread, how many days would it take for the patch to cover half of the loaf of bread?
\vspace{0.25cm}

\noindent Have you seen any of the last 3 word problems before?\vspace{-0.25cm}
\begin{itemize}
    \item Yes
    \vspace{-0.25cm}
    \item Maybe
    \vspace{-0.25cm}
    \item No
\end{itemize}

\subsection{Berlin Numeracy Test}
\noindent Imagine we are throwing a five-sided die 50 times. On average, out of these 50 throws how many times would this five-sided die show an odd number (1, 3 or 5)? \\
---out of 50 throws.
\vspace{0.25cm}

\noindent Out of 1,000 people in a small town 500 are members of a choir. Out of these 500 members in the choir 100 are men. Out of the 500 inhabitants that are not in the choir 300 are men. What is the probability that a randomly drawn man is a member of the choir? (please indicate the probability in percent). \\
---\%
\vspace{0.25cm}

\noindent Imagine we are throwing a loaded die (6 sides). The probability that the die shows a 6 is twice as high as the probability of each of the other numbers. On average, out of these 70 throws, how many times would the die show the number 6? 
--- out of 70 throws.
\vspace{0.25cm}

\noindent In a forest 20\% of mushrooms are red, 50\% brown and 30\% white. A red mushroom is poisonous with a probability of 20\%. A mushroom that is not red is poisonous with a probability of 5\%. What is the probability that a poisonous mushroom in the forest is red? 
---\%

\subsection{Political Knowledge}
\noindent In the next part of this study, you will be asked several factual questions about politics and public policy. Many people don’t know the answers to these questions, but it is helpful for us if you answer, even if you’re not sure what the correct answer is. We encourage you to take a guess on every question. Please just give your best guess. 

\noindent **Do not look up the answers in a book or on the Internet.** 

\noindent You will be given 10 seconds to respond to each question before the survey will advance.
\vspace{0.25cm}

\noindent Whose responsibility is it to decide if a law is constitutional or not?\vspace{-0.25cm}
\begin{itemize}
    \item The President
    \vspace{-0.25cm}
    \item Congress
    \vspace{-0.25cm}
    \item The Supreme Court
\end{itemize}

\noindent Whose responsibility is it to nominate judges to Federal Courts?\vspace{-0.25cm}
\begin{itemize}
    \item The President
    \vspace{-0.25cm}
    \item Congress
    \vspace{-0.25cm}
    \item The Supreme Court
\end{itemize}

\noindent Who is the leader of the Labour Party of Great Britain? Is it:\vspace{-0.25cm}
\begin{itemize}
    \item Theresa May
    \vspace{-0.25cm}
    \item Keir Starmer
    \vspace{-0.25cm}
    \item Tony Hayward
    \vspace{-0.25cm}
    \item Boris Johnson
\end{itemize}

\noindent Do you know what job or political office is currently held by Nancy Pelosi? Is it:\vspace{-0.25cm}
\begin{itemize}
    \item Speaker of the House
    \vspace{-0.25cm}
    \item Treasury Secretary
    \vspace{-0.25cm}
    \item Senate Majority Leader
    \vspace{-0.25cm}
    \item Justice of The Supreme Court
    \vspace{-0.25cm}
    \item Governor of New Mexico
\end{itemize}

\noindent Do you know what job or political office is currently held by Steve Mnuchin? Is it:\vspace{-0.25cm}
\begin{itemize}
    \item Attorney General
    \vspace{-0.25cm}
    \item Justice of The Supreme Court
    \vspace{-0.25cm}
    \item Treasury Secretary
    \vspace{-0.25cm}
    \item House Republican Leader
    \vspace{-0.25cm}
    \item Secretary of State
\end{itemize}

\subsection{Attitudes Towards Science}
\noindent We are also interested in your opinions about scientific research.
\vspace{0.25cm}

\noindent Because of science and technology, there will be more opportunities for the next generation.
\vspace{-0.25cm}
\begin{itemize}
    \item Strongly disagree
    \vspace{-0.25cm}
    \item Disagree
    \vspace{-0.25cm}
    \item Agree
    \vspace{-0.25cm}
    \item Strongly agree
\end{itemize}

\noindent Science makes our way of life change too fast.
\vspace{-0.25cm}
\begin{itemize}
    \item Strongly disagree
    \vspace{-0.25cm}
    \item Disagree
    \vspace{-0.25cm}
    \item Agree
    \vspace{-0.25cm}
    \item Strongly agree
\end{itemize}

\noindent Even if it brings no immediate benefits, scientific research that advances the frontiers of knowledge is necessary and should be supported by the federal government.
\vspace{-0.25cm}
\begin{itemize}
    \item Strongly disagree
    \vspace{-0.25cm}
    \item Disagree
    \vspace{-0.25cm}
    \item Agree
    \vspace{-0.25cm}
    \item Strongly agree
\end{itemize}

\noindent Would you say that, on balance, the benefits of scientific research have outweighed the harmful results?
\vspace{-0.25cm}
\begin{itemize}
    \item Strongly disagree
    \vspace{-0.25cm}
    \item Disagree
    \vspace{-0.25cm}
    \item Agree
    \vspace{-0.25cm}
    \item Strongly agree
\end{itemize}

\subsection{Demographics}
\noindent What is your age?
\vspace{0.25cm}

\noindent What is your gender?
\vspace{-0.25cm}
\begin{itemize}
    \item Male
    \vspace{-0.25cm}
    \item Female
    \vspace{-0.25cm}
    \item Other
\end{itemize}

\noindent What is the highest level of school you have completed or the highest degree you have received? 
\vspace{-0.25cm}
\begin{itemize}
    \item Less than high school degree
    \vspace{-0.25cm}
    \item High school graduate (high school diploma or equivalent including GED)
    \vspace{-0.25cm}
    \item Some college but no degree
    \vspace{-0.25cm}
    \item Associate degree in college (2-year)
    \vspace{-0.25cm}
    \item Bachelor's degree in college (4-year)
    \vspace{-0.25cm}
    \item Master's degree
    \vspace{-0.25cm}
    \item Doctoral degree
    \vspace{-0.25cm}
    \item Professional degree (JD, MD)
\end{itemize}

\noindent Information about income is very important to understand. Would you please give your best guess?\\
Please indicate the answer that includes your entire household income in (previous year) before taxes.
\vspace{-0.25cm}
\begin{itemize}
    \item Less than \$10,000
    \vspace{-0.25cm}
    \item \$10,000 to \$19,999
    \vspace{-0.25cm}
    \item \$20,000 to \$29,999
    \vspace{-0.25cm}
    \item \$30,000 to \$39,999
    \vspace{-0.25cm}
    \item \$40,000 to \$49,999
    \vspace{-0.25cm}
    \item \$50,000 to \$59,999
    \vspace{-0.25cm}
    \item \$60,000 to \$69,999
    \vspace{-0.25cm}
    \item \$70,000 to \$79,999
    \vspace{-0.25cm}
    \item \$80,000 to \$89,999
    \vspace{-0.25cm}
    \item \$90,000 to \$99,999
    \vspace{-0.25cm}
    \item \$100,000 to \$149,999
    \vspace{-0.25cm}
    \item \$150,000 or more
\end{itemize}

\noindent Are you fluent in English?
\vspace{-0.25cm}
\begin{itemize}
    \item Yes
    \vspace{-0.25cm}
    \item No
\end{itemize}

\noindent Please choose whichever ethnicity that you identify with (you may choose more than one option):
\vspace{-0.25cm}
\begin{itemize}
    \item White/Caucasian
    \vspace{-0.25cm}
    \item Black or African American
    \vspace{-0.25cm}
    \item American Indian or Alaska Native
    \vspace{-0.25cm}
    \item Asian
    \vspace{-0.25cm}
    \item Native Hawaiian or Pacific Islander
    \vspace{-0.25cm}
    \item Other
\end{itemize}

\noindent Which of the following best describes your political position?
\vspace{-0.25cm}
\begin{itemize}
    \item Democrat
    \vspace{-0.25cm}
    \item Republican
    \vspace{-0.25cm}
    \item Independent
    \vspace{-0.25cm}
    \item Other (specify)
\end{itemize}

\noindent Which of the following best describes your political preference?
\vspace{-0.25cm}
\begin{itemize}
    \item Strongly Democratic
    \vspace{-0.25cm}
    \item Democratic
    \vspace{-0.25cm}
    \item Lean Democratic
    \vspace{-0.25cm}
    \item Lean Republican
     \vspace{-0.25cm}
    \item Republican
    \vspace{-0.25cm}
    \item Strongly Republican
\end{itemize}

\noindent We are also interested in your *past* political preferences.\\
Please select any of the following preferences that have *ever* applied to you in the past. (You may select as many as you like)
\vspace{-0.25cm}
\begin{itemize}
    \item Strongly Democratic
    \vspace{-0.25cm}
    \item Democratic
    \vspace{-0.25cm}
    \item Lean Democratic
    \vspace{-0.25cm}
    \item Lean Republican
     \vspace{-0.25cm}
    \item Republican
    \vspace{-0.25cm}
    \item Strongly Republican
\end{itemize}

\noindent On social issues I am:
\vspace{-0.25cm}
\begin{itemize}
    \item Strongly Liberal
    \vspace{-0.25cm}
    \item Somewhat Liberal
    \vspace{-0.25cm}
    \item Moderate
    \vspace{-0.25cm}
    \item Somewhat Conservative
     \vspace{-0.25cm}
    \item Strongly Conservative
\end{itemize}

\noindent On economic issues I am:
\vspace{-0.25cm}
\begin{itemize}
    \item Strongly Liberal
    \vspace{-0.25cm}
    \item Somewhat Liberal
    \vspace{-0.25cm}
    \item Moderate
    \vspace{-0.25cm}
    \item Somewhat Conservative
     \vspace{-0.25cm}
    \item Strongly Conservative
\end{itemize}

\noindent How have your politics changed since you were young (if at all)? 
\vspace{-0.25cm}
\begin{itemize}
    \item I am more strongly liberal
    \vspace{-0.25cm}
    \item I am somewhat more strongly liberal
    \vspace{-0.25cm}
    \item No change
    \vspace{-0.25cm}
    \item I am somewhat more strongly conservative
     \vspace{-0.25cm}
    \item I am more strongly conservative
\end{itemize}

\noindent Who did you vote for in the 2016 Presidential Election (if anyone)?\\
Reminder: This survey is anonymous.
\vspace{-0.25cm}
\begin{itemize}
    \item Hillary Clinton
    \vspace{-0.25cm}
    \item Donald Trump
    \vspace{-0.25cm}
    \item Other candidate (such as Jill Stein or Gary Johnson)
    \vspace{-0.25cm}
    \item I did not vote for reasons outside of my control
     \vspace{-0.25cm}
    \item I did not vote, but I could have
    \vspace{-0.25cm}
    \item I did not vote out of protest
\end{itemize}

\noindent Who did you vote for in the 2018 Congressional Election (if anyone)? \\
Reminder: This survey is anonymous.
\vspace{-0.25cm}
\begin{itemize}
    \item The Democratic Party candidate in my district
    \vspace{-0.25cm}
    \item The Republican Party candidate in my district
    \vspace{-0.25cm}
    \item Other candidate
    \vspace{-0.25cm}
    \item I did not vote for reasons outside of my control
     \vspace{-0.25cm}
    \item I did not vote, but I could have
    \vspace{-0.25cm}
    \item I did not vote out of protest
\end{itemize}

\noindent What percentage of your social circle (friends and family) do you think votes like yourself?  \\
(For example: If you vote Democratic, what percentage also votes Democratic? / If you vote Republican, what percentage also votes Republican?)
\vspace{0.25cm}

\noindent ``My political attitudes and beliefs are an important reflection of who I am''
\vspace{-0.25cm}
\begin{itemize}
    \item Strongly disagree
    \vspace{-0.25cm}
    \item Disagree
    \vspace{-0.25cm}
    \item Somewhat disagree
    \vspace{-0.25cm}
    \item Neither agree nor disagree
    \vspace{-0.25cm}
    \item Somewhat agree
    \vspace{-0.25cm}
    \item Agree
    \vspace{-0.25cm}
    \item Strongly agree
\end{itemize}

\noindent ``In general, my political attitudes and beliefs are an important part of my self-image''
\vspace{-0.25cm}
\begin{itemize}
    \item Strongly disagree
    \vspace{-0.25cm}
    \item Disagree
    \vspace{-0.25cm}
    \item Somewhat disagree
    \vspace{-0.25cm}
    \item Neither agree nor disagree
    \vspace{-0.25cm}
    \item Somewhat agree
    \vspace{-0.25cm}
    \item Agree
    \vspace{-0.25cm}
    \item Strongly agree
\end{itemize}

\noindent Generally speaking, how interested are you in politics and public affairs?
\vspace{-0.25cm}
\begin{itemize}
    \item Not at all
    \vspace{-0.25cm}
    \item Not very
    \vspace{-0.25cm}
    \item Somewhat
    \vspace{-0.25cm}
    \item Extremely
\end{itemize}

\noindent Generally speaking, how often do you pay attention to information about politics and public affairs?
\vspace{-0.25cm}
\begin{itemize}
    \item Not at all
    \vspace{-0.25cm}
    \item Sometimes
    \vspace{-0.25cm}
    \item About half of the time
    \vspace{-0.25cm}
    \item Most of the time
    \vspace{-0.25cm}
    \item All of the time
\end{itemize}

\noindent Some people think that by criticizing leaders, news organizations keep political leaders from doing their job. Others think that such criticism is worth it because it keeps political leaders from doing things that should not be done. Which position is closer to your opinion?
\vspace{-0.25cm}
\begin{itemize}
    \item Criticism from news organizations keeps political leaders from doing their job.
    \vspace{-0.25cm}
    \item Criticism from news organizations keeps political leaders from doing things that should not be done.
\end{itemize}

\noindent In presenting the news dealing with political and social issues, do you think that news organizations deal fairly with all sides, or do they tend to favor one side?
\vspace{-0.25cm}
\begin{itemize}
    \item News organizations tend to deal fairly with all sides.
    \vspace{-0.25cm}
    \item News organizations tend to favor one side.
\end{itemize}

\noindent To what extent do you trust the information that comes from the following?
\vspace{-0.25cm}
\begin{itemize}
    \item National news organizations
    \vspace{-0.25cm}
    \item Local news organizations
    \vspace{-0.25cm}
    \item Friends and family
    \vspace{-0.25cm}
    \item Social networking sites (e.g., Facebook, Twitter)
\end{itemize}
[The responses for each of the above items in the previous question are: None at all, A little, A moderate amount, A lot, A great deal.]
\vspace{0.25cm}

\noindent How strongly do you believe in the existence of a God or Gods?
\vspace{-0.25cm}
\begin{itemize}
    \item 1 - Not at all
    \vspace{-0.25cm}
    \item 2
    \vspace{-0.25cm}
    \item 3
    \vspace{-0.25cm}
    \item 4
    \vspace{-0.25cm}
    \item 5
    \vspace{-0.25cm}
    \item 6
    \vspace{-0.25cm}
    \item 7 - Very much
\end{itemize}

\noindent Which of the following best describes your current stance toward God (or gods)?
\vspace{-0.25cm}
\begin{itemize}
    \item I believe in God
    \vspace{-0.25cm}
    \item I don't really take a stance on God
    \vspace{-0.25cm}
    \item I don't know whether or not God exists
    \vspace{-0.25cm}
    \item I don't believe in God
\end{itemize}

\noindent We are interested in what people *used to* believe about God (or gods).\\
Thinking about your past, which of the following categories would you have fit into (however loosely) at some point in your life (excluding childhood, of course)? \\
(please choose all that applied to you at any point in your past)
\vspace{-0.25cm}
\begin{itemize}
    \item I believed in God
    \vspace{-0.25cm}
    \item I didn't really take a stance on God
    \vspace{-0.25cm}
    \item I didn't know whether or not God exists
    \vspace{-0.25cm}
    \item I didn't believe in God
\end{itemize}

\noindent How have your religious beliefs changed since you were young (if at all)?
\vspace{-0.25cm}
\begin{itemize}
    \item I am more strongly non-religious
    \vspace{-0.25cm}
    \item I am somewhat more strongly non-religious
    \vspace{-0.25cm}
    \item No change
    \vspace{-0.25cm}
    \item I am somewhat more strongly religious
    \vspace{-0.25cm}
    \item I am more strongly religious
\end{itemize}

\noindent Please enter the ZIP code for your primary residence.\\
Reminder: This survey is anonymous.  
\vspace{0.25cm}

\noindent Do you have any comments about our survey?